# Crossover between re-nucleation and dendritic growth in electrodeposition without supporting electrolyte

C. Kharbachi, T. Tzedakis, F. Chauvet[*]

*Laboratoire de Génie Chimique, Université de Toulouse, CNRS, INPT, UPS, Toulouse, France*

This work focuses on the microstructure of metallic deposits formed by galvanostatic electrodeposition inside a Hele-Shaw cell without both supporting electrolyte and flow. For a low applied current density $j$, the deposit grows under the form of ramified branches which form a bidimensional fractal pattern. As shown by Fleury (V. Fleury, Nature, 390, 145-148, 1997), these branches are composed of small metallic crystals. This microstructure is built up by a re-nucleation process induced by the dynamics of a space charge region (non-electrically neutral solution) ahead of the growth front. When increasing $j$ the crystal size decreases whereas the nucleation frequency increases. These latter tendencies are reversed for high $j$ when, as experimentally observed, dendrites are formed instead of ramified branches. There must be a transition between the nucleation/growth regime (ramified branches) and the pure growth regime (dendrites). This transition is examined experimentally by carefully observing the branch microstructure by Scanning Electron Microcopy (SEM) and by varying both $j$ and the electrolyte concentration $c_0$. For copper and silver branches, when $j$ is lower than a critical current density $j_c$ (concentration-dependent), the branches are composed only of non-dendritic crystals. Whereas, when $j > j_c$, dendritic crystals are observed and they become the main kind of crystals constituting the branches for higher $j$. These observations show that the morphological transition on the pattern scale, between ramified branches and dendrites, originates from a morphological transition on the scale of the crystals constituting the branches. This latter is considered theoretically by analyzing the shape stability of the growing crystals. The Mullins & Sekerka model (shape stability of a spherical particle growing by diffusion) disagrees with these observations by predicting that the crystals are always unstable. It is proposed that the space charge layer, surrounding the growing crystals, induces a stabilizing effect.

# I. INTRODUCTION

The electrodeposition of a metal is known to be able to lead to the formation of metal ramified structures commonly called *dendrites*. The controlled formation of these ramified structures in Hele-Shaw cells leads to the fast formation of bidimensional patterns visible at the naked eye. The formation of such structures is primary explains by the fact that no supporting electrolyte is used. As a consequence, to avoid the separation of the charge carriers in the solution (anions and cations of the dissolved metal salt), i.e. the deviation from electroneutrality, the growth front must move towards the anode at the same velocity as that of the anions [1–4]. As a consequence, since this velocity is typically high (~10 µm/s), the deposit density is very low that leads to ramified structures [1,5]. During the 80's and 90's, the electrochemical growth of these structures was used as a model situation for theoretical studies on pattern formation in connection with Diffusion Limited Aggregation (DLA) patterns [6–13]. From these works it appeared that, depending on the operating parameters (imposed electric current, concentration of metal salt, etc.) and on the nature of the metal, either ramified branches or dendrites are formed on the macroscopic scale (visible at the naked eye) [8,14,15].

The ramified branches are fractal on a given range of length scales [16]. This comes from the fact that the growth front is unstable for a wide range of wavenumbers. The growth process is mainly diffusion-limited, without stabilizing effects (surface energy and kinetics) at the corresponding length scales [9]. The pattern of ramified branches can be of two types: fractal-like and Dense Branching (DB). A fractal-like pattern is obtained in the limit of low growth velocities (low applied currents) [16]. Under this condition, the growth process is then close to the Laplacian growth model giving rise to a pure fractal deposit, with an infinite upper cut-off length scale, i.e. a DLA pattern [9,17]. The DB patterns are obtained for higher and finite growth velocities. The branches then grow behind an almost flat growth front. The DB patterns are also fractal but on a limited range of length scales with a finite upper bound due to the stabilization of the growth front [18]. As resumed by Léger et al. 2000 [5], the transition between fractal-like and DB patterns is still not well understood. Indeed, this is difficult to know if this transition is intrinsically related to the diffusion because of the presence of interfering effects such as several kinds of convection (natural convection [12,19], electroconvection [20] and electro-osmosis [12]) or the electrical resistance of the branches themselves [21].

Dendritic deposits are typically observed for even higher applied currents than for DB patterns and mainly with zinc [7,8]. Note that here, the term dendrite is used to name single crystals which have been subjected to an out-of-equilibrium growth with anisotropy effects (surface energy and/or kinetics). This results in crystal shapes with well-ordered geometry, such as a main arm with different levels of side branching oriented with a constant angle. This differs from the random shape of the ramified branches. The structures of both ramified branches and dendrites are sketched in Fig. 1. Grier et al. 1986 [7] explain the appearance of the dendrites (on the macroscopic scale) by an exacerbated difference in growth rate, as a function of the orientation, when the current is high. In other terms, this corresponds to the out-of-equilibrium growth of a single anisotropic crystal. Obviously, the dendrite shape depends on the crystal structure of the metal, as noted by Grier et al. 1986 [7] by comparing copper and zinc deposits on the macroscopic scale.

There is a significant difference in microstructure between dendritic deposits and ramified branches, as revealed by X-ray diffraction [7]. The X-ray diffraction pattern is, as expected, anisotropic for dendrites (single crystals) but isotropic for ramified branches. This suggests a polycrystalline structure for ramified branches.

Very little work has focused on the micro/nano structure of the ramified branches. One reason is the difficult recovery of these fragile branches without damaging them. Using a cell, in which the inner bottom wall is covered with a non-percolating gold coating (metallization by vapor deposition), Fleury [22] succeeded in forming copper ramified branches that adhere to this wall. SEM observations of these plates show that the branches consist of small metallic (non-dendritic) crystals whose size could be lower than 100 nm. To date, Fleury proposes the most advanced interpretation and modeling for the building of this structure during branch growth: in order for the average growth velocity to match the velocity of anions, new crystals have to periodically nucleate and grow at the top of the branches. This principle is based on the fact that, assuming a given crystal grows at a constant rate, after a given duration, the local growth velocity is lower than the velocity of the anions [22]. This leads, in the vicinity of the metal surface, to the formation of a space charge region where the electroneutrality is not respected [2]. Furthermore, at the surface of the growing metal, the electrolyte depletion induces the increase of both the electric field $E_s$ and the cell tension [23,24]. When $E_s$ exceeds a given threshold, a new crystal nucleates and grows and the cycle restarts [22]. Fleury showed that the crystal size decreases with the current density. He also provided a relation between the average growth velocity $v_g$, the crystal size $R_g$ and the crystal growth time $T$ : $v_g = 2R_g/T$. This latter relation also gives access to the nucleation frequency $T^{-1} = v_g/(2R_g)$. This law shows that $T^{-1}$ increases with the current density, the latter causing $v_g$ to increase and $R_g$ to decrease, Fig. 1. However, this evolution of the nucleation frequency is not consistent with the formation of dendrites in the limit of high current densities. Indeed, the dendrites are built only by growth without re-nucleation events. Consequently, $T^{-1}$ must vanish at high current densities. There must be a transition from the nucleation/growth regime (ramified branches) to the growth regime (dendrites) as sketched in Fig. 1. In this paper, this transition is studied experimentally and theoretically on the scale of the branch microstructure.

Concerning the experimental part, we use the same approach as Fleury, by performing SEM observations of the branches. We do not wish our analysis to be dependent on the surface state of the cell wall. We are interested in the intrinsic transition between the two regimes. Consequently, the inner wall of the cells used in this study is not activated as in the work of Fleury [22]; in an initial study, we show how both the pattern and the microstructure are affected by the surface state of the cell wall. A specific method, based on the freezing of the cell, is used to recover the branches without damage. Applied current densities, higher than in Fleury's work, are used to cover both regimes. The main part of the study concerns the growth of copper branches, but some experiments are performed with silver, to investigate the effect of the nature of the deposited metal.

As for the theoretical part, the onset of dendritic growth is analyzed by considering the required shape instability of the growing material. This is done by adapting the Mullins & Sekerka shape stability model [25] to the electrochemical situation. How the obtained instability threshold depends on the operating parameters (current density $j$ and metal salt concentration $c_0$) is discussed and compared to experiment results.

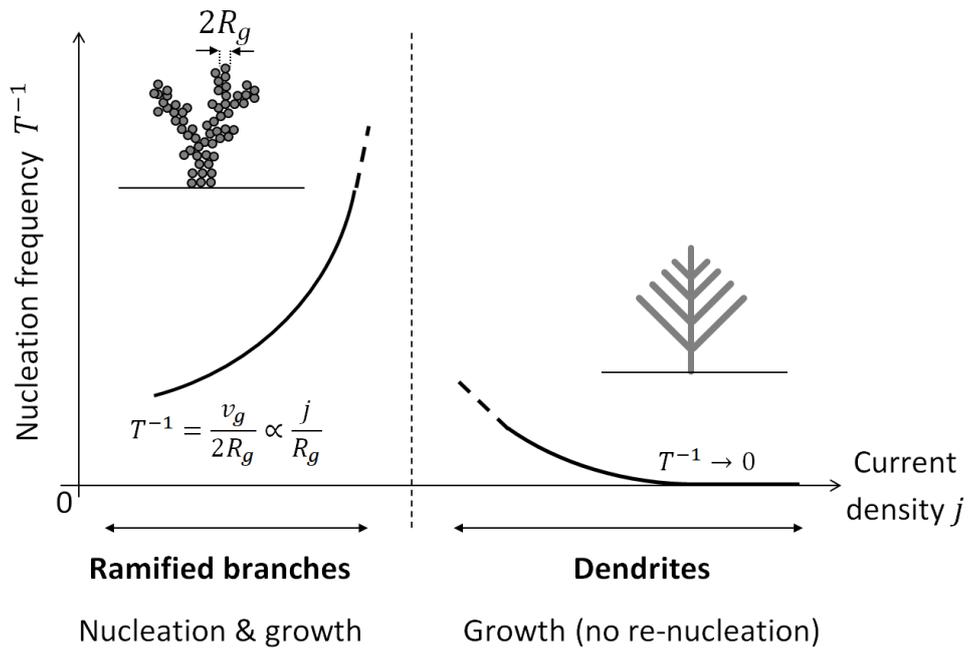

FIG. 1. Illustration of the dependence of the nucleation frequency $T^{-1}$ on the applied current density $j$, when the growth regime changes from re-nucleation process (ramified branches) to pure growth without re-nucleation process (dendrites).

## II. EXPERIMENTAL SET-UP AND METHODS

### A. Chemicals

The aqueous metal salt solutions (0.1 to 0.75 M) are made by dissolving either copper (II) sulfate pentahydrate (>98%, Sigma Aldrich) or silver nitrate (>99%, Acros Organics) in deionized water (18.2 MΩ.cm). Prior using the solutions, the dissolved oxygen is removed by bubbling nitrogen for 10 minutes (~5 mL). The deaerated solutions are then collected by a gastight syringe (Hamilton 1 mL, 1001LT) and injected into the Hele-Shaw cell.

### B. Experimental set-up

The experimental set-up (Fig. 2) is made from two microscope glass plates (76×56×1 mm) which sandwich the electrodes of thickness $e = 50$ μm. The depth of the Hele-Shaw cell therefore corresponds to the thickness of the electrodes. The length and width $L$ of the electrolytic compartment are respectively 75 mm and 15 mm. The same metal element was used for both electrodes (metallic form) and for the electrolyte (dissolved metal salt). The purity of the metal electrodes is 99.9% for copper and 99.9% for silver (Goodfellow). To avoid leakage between the electrodes and the glass plates, the inner sides of the glass plates are covered by a transparent laboratory parafilm. The effect of the functionalization of the inner cell walls is investigated by metallizing the parafilm by gold vapor deposition. The obtained gold coating is non-percolating (as in Fleury's work) as confirmed by resistivity measurements.

The electrodes are polished with polishing paper and cleaned with ethanol and then flushed with deionized water before each use. The two remaining opening sides of the cell are closed by applying a reusable adhesive paste (UHU patafix). Fluidic connections are made using nanoport connectors (Idex-hs) on holes drilled in the glass (Dremel diamond wick). This assembly is kept horizontal in a mount made of Plexiglas. The cell is then filled with the electrolyte solution previously collected with the syringe. After each experiment, the device is opened, cleaned and re-assembled for the next experiment.

The galvanostatic electrolyses are performed using either a potentiostat (Autolab PGSTAT100N) or a current generator (TDK-Lambda Gen2400W) to impose an electric current ranging from 1.25 to 20 mA ($j$ is in the range 33 to 266 mA/cm$^2$) between the electrodes, regardless of the electrical resistance of the system. All experiments are performed under ambient conditions (~20°C).

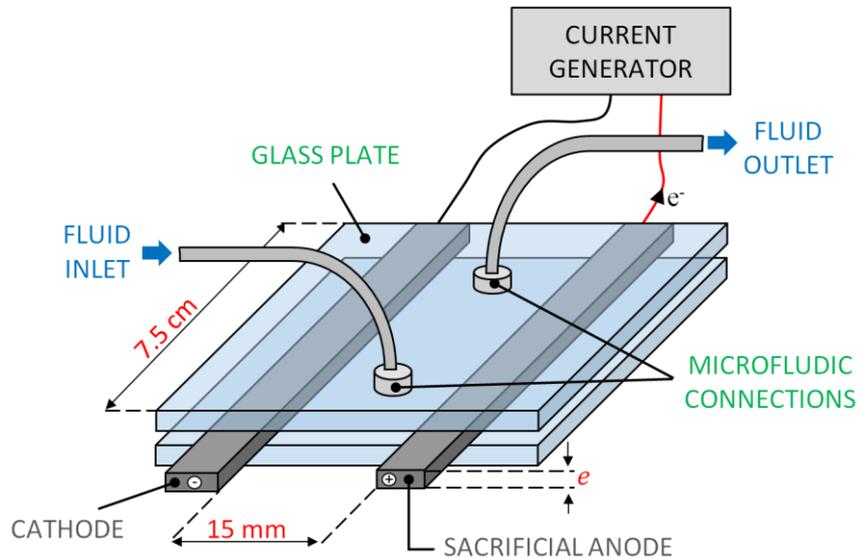

FIG.2. Sketch of the experimental set-up

### C. Branch recovery method

Just after the electrochemical formation of the metallic branches, electrolysis is stopped, and a cooling spray, inducing a temperature decrease of ~50°C (RS components 846-682), is applied to the top glass plate of the device to freeze the liquid electrolyte. This step allows the inlet fluidic connection to be opened without disturbing the highly fragile electrodeposit, in order to connect a syringe of deionized and deaerated water. After, the cell is left to unfreeze and return to room temperature where a flow is applied for 30 minutes to gently flush the metallic branches (~20 μL/min) using a syringe pump (Harvard Apparatus PHD Ultra 70-3006). This allows the removal of almost all the electrolyte thus avoiding any eventual crystallization in samples examined by SEM.

The cell is then frozen again, with the cooling spray, to disconnect the two fluidic connections before putting the entire system in a freezer for ~30 min. Next, the cell is opened quickly on an ice bed and the branch pattern is transferred to an adhesive carbon tape where it is left free to dry under ambient conditions.

This method enables the recovery of undisturbed branches and their analysis by SEM on a wide range of length scales (from ~1 mm up to ~10 nm), as shown in Fig. 3.

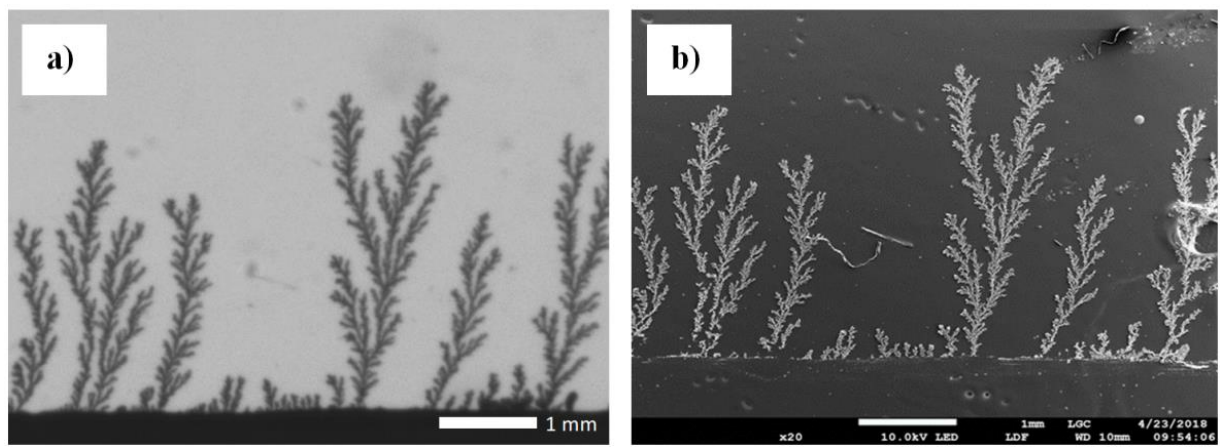

FIG. 3. a) Optical image of ramified copper branches taken during a galvanostatic electrolysis in the Hele-Shaw cell (elapsed time = 220 s, $c_0$ = 0.5 M and $j$ = 133 mA/cm$^2$), b) SEM image taken after the recovery of the deposit.

### D. Visualization of the branch pattern

The growth of the branches is visualized by transmission using a LED panel, placed below the device, and a camera PCO pixelfly, associated with a 105 mm macro lens, facing the top glass plate. The acquisition frequency is 1 image/s unless another value is specified. The Fiji software and python scripts are used for image processing.

### E. Characterization of branch microstructure

All the branch samples, obtained under various operating conditions, are metalized by vapor deposition of a layer of ~10 nm of gold or platinum (duration = 60 s, vacuum = 10$^{-1}$ mbar) before being observed by SEM using a MEB-FEG JEOL JSM 7800F Prime - EDS or a JEOL JSM 7100F TTLS.

# III. MICROSTRUCTURE OF THE RAMIFIED BRANCHES

The ramified branches are observed by SEM. Several experiments are carried out varying $j$ and $c_0$. Additionally, this is done with and without metallization of the cell wall to investigate its effect on the branch microstructure. The corresponding results are first given and compared in the following sub-section. Then, the results obtained without metallization are given and analyzed in the subsequent sub-section.

## A. Effect of the metallization of the cell wall

The effect of the metallization of the cell wall is shown by comparing the microstructures of copper branches (applied current density $j = 133$ mA/cm$^2$, electrolyte concentration $c_0 = 0.5$ M) with and without cell wall metallization, Fig. 4. With metallization, as already reported in previous works [26,27], the branches adhere to the cell wall; the deposit is thus easily recovered, flushed and analyzed by SEM (as in the Fleury's work [22]). The corresponding SEM images, Fig. 4a-b, show that the branches have the form of "compact tongues" composed of small crystals (equivalent diameters range from ~100 to ~600 nm). Without metallization, the branches do not adhere to the cell wall; their recovery and SEM analyses are achieved from the method described in section II. In this case, the branch microstructure is different exhibiting an expanded structure. These branches are composed of small non-dendritic and dendritic crystals, Fig. 4c-d. The same differences, with and without metallization, have been observed when varying the values of the operating parameters ($j$ and $c_0$) in the investigated ranges.

These results show a strong influence of the interaction between the cell wall and the electrodeposited metal on the branch microstructure. In the case of adherent branches (with metallization), the growth phenomenon is complex since the microstructure should clearly depend on the properties of the metallized film (nature of the metal, size distribution, average thickness, etc.); these properties must therefore be considered for a full description of the branch growth as initiated in previous works [26,27].

In this work, it is desired to study the still poorly explored intrinsic transition between ramified branches and dendritic deposits (on the scale of the microstructure) regardless of the surface state of the cell wall.

Consequently, the study is performed with non-adherent branches obtained without metallization of the cell wall.

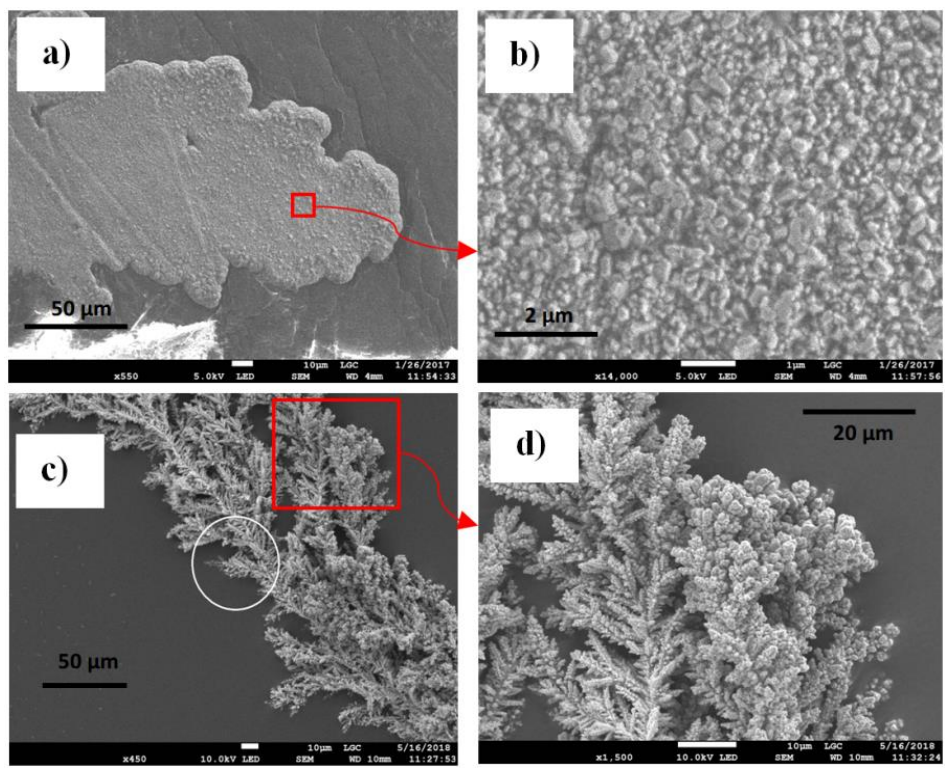

FIG. 4. SEM images of copper branches obtained with gold metallization of cell walls, at a magnification of ×550 in a) and ×14000 in b), and without metallization (with cell walls made of bare parafilm), at magnification of ×450 in c) and ×15000 in d); $j = 133$ mA/cm$^2$ and $c_0 = 0.5$ M.

### B. Microstructure of non-adherent branches

The recovery method allows the analysis of the microstructure of a given non-adherent branch, as shown in Fig. 5 for a concentration of 0.5 M and two applied current densities of 66 and 133 mA/cm$^2$. At a low current density of 66 mA/cm$^2$, the copper branches look like assemblies of non-dendritic crystals similar to the branch structures obtained by Fleury, Fig. 5a-d. At a higher current density, of 133 mA/cm$^2$ (Fig. 5e-h), dendritic shapes are visible in the structure of the branches. In the following, such a visual detection of the presence of dendrites is indicated by the sketch of a dendrite in the corresponding SEM images. Both single dendrites (Fig. 5f) and dendrites on which non-dendritic crystals have nucleated and grown (Fig. 5g) are observed. As shown in Fig. 6, an increase in the current density, up to 266 mA/cm$^2$ ($c_0$ = 0.5 M), reinforces the presence of dendrites compared to non-dendritic crystals, even if on the pattern scale, no dendritic structure is observed (Fig. 5e and Fig. 6a). Experiments carried out with other concentration, $c_0$ = 0.25 M and 0.75 M, also lead to the observation of this transition between non-dendritic and dendritic copper crystals, as shown in Fig. 7. However, the transition appears later in terms of $j$ when $c_0$ is increased.

To avoid any subjective bias, in in the highlighting of the transition between non-dendritic and dendritic crystals, the presence of dendritic crystals is determined from image processing based on the detection of preferential orientations (revealing dendrites) in the SEM images. This is performed using the plugin directionality of the Fiji software. This plugin computes an indicator of preferential orientations in an image. Here, we chose the method based on the Fourier Transform, for which the indicator corresponds to the radially averaged FFT spectra. For an isotropic structure the plot of the indicator, as a function of the orientation, is flat. For an anisotropic structure the preferential orientations are highlighted by the presence of peaks on the plot of the indicator. The application of this treatment to the SEM images is shown in Fig. 8 for two applied current densities (66 and 133 mA/cm$^2$) and $c_0$ = 0.5 M. The presence of dendrites in the SEM image (Fig. 8c) is well indicated by peaks in the corresponding directionality plot. For the SEM image without apparent dendrites (Fig. 8b), the directionality plot is flat. A descriptor of the presence of dendrites can be defined as the ratio between the standard deviation and the mean value of a given directionality plot; low values of the descriptor indicate the absence of dendrites whereas high values indicate their presence. In Fig. 9, this dendrite descriptor is plotted as a function of $j$ and for all the concentrations (except 0.1 M for which an unaltered deposit could not be recovered). Note that this image processing is performed on images at the same magnification x15,000. Despite a certain dispersion of the obtained values, we observe an increase of the dendrite descriptor with $j$ for all cases. From these plots, a critical current density $j_c$, corresponding to the appearance of dendrites, can be defined. This is done by fitting the data with a erf function using the same least mean square process for the three concentrations; we use the python function scipy.optimize.curve_fit with bounds : [0-0.3] for the amplitude, [0-200] for $j_c$ and [0-500] for the wave width. The corresponding fits are shown in Fig. 9 as solid lines. The obtained critical current densities $j_c$ are plotted as a function of $c_0$ in the stability diagram shown in Fig. 10. The stability curve $j_c = f(c_0)$ (black line) separates non-dendritic and dendritic regimes. As expected from the analysis of the SEM images (Fig. 7), the critical current density $j_c$ is found to increase with the concentration $c_0$. Note that the values of $j_c$ depend on both the method used for the quantification of the dendrites and the function used for data fitting. Therefore, the as-obtained values of $j_c$ are method-dependent and they cannot be assimilated to intrinsic physical properties. Nevertheless, they provide an estimate of the actual threshold and most importantly, they indicate how the transition is affected by the metal salt concentration $c_0$.

The observed microstructures are always the same whatever the location along the branches (Fig. 5 and 6). This shows that the same microstructure is continuously formed during the branch growth. As a consequence, it is deduced that the dendrites arise from shape instability of some initially non-dendritic crystals, during their growth at the top of the branches. As sketched in Fig. 11, for $j > j_c$, some crystals are unstable and they grow faster than the stable crystals. Whereas for $j < j_c$, all the crystals are stable. For $j > j_c$, both dendritic and non-dendritic crystals are observed but the amount of dendritic crystals, compared to non-dendritic crystals, seems to increase with $j$, Fig. 6. However, note that even at high $j$, the re-nucleation process is still ongoing (for copper). This probably allows the changes in growth direction of the branches which are still ramified at high $j$ (Fig. 6a). On the pattern scale, since the deposit is fractal, the growth direction of a ramified branch is random. On the contrary, for a dendrite, the growth direction would be fixed and defined by both the crystal structure and the initial growth orientation.

Note that the dendritic growth is not initiated from a shape instability of the initial flat electrode surface because in this case we would see only dendrites even at the pattern scale.

Some experiments were carried out by replacing the copper electrodes by silver electrodes and the copper sulfate solution by a silver nitrate solution at a concentration of 0.25 M. Fig. 12 shows the silver branch microstructure obtained for four applied current densities (33, 66, 133 and 266 mA/cm$^2$). At the lowest $j$ (33 mA/cm$^2$), compared to the other cases, numerous small crystals are visible; note that the magnification is similar for each case. This microstructure must necessarily come from a re-nucleation process as for the copper at low $j$. For the other $j$ (66-266 mA/cm$^2$), the deposit consists of bigger crystals which are elongated and sharp and can be considered as dendrites even without side-branching [9]. The specific shape of these dendrites is clearly related to an anisotropy effect of the silver which differs from that of copper. At high $j$, the formation of the silver deposit is mainly ensured by the crystal growth rather than the re-nucleation process. We conclude that the same transition as for copper is therefore observed for silver.

To sum up, the expected transition between the re-nucleation regime (ramified branches) and the growth regime (dendrites) is observed on the scale of the crystals constituting the branches. The critical current density $j_c$ increases with the concentration $c_0$. This experimentally observed transition is considered theoretically in the following section.

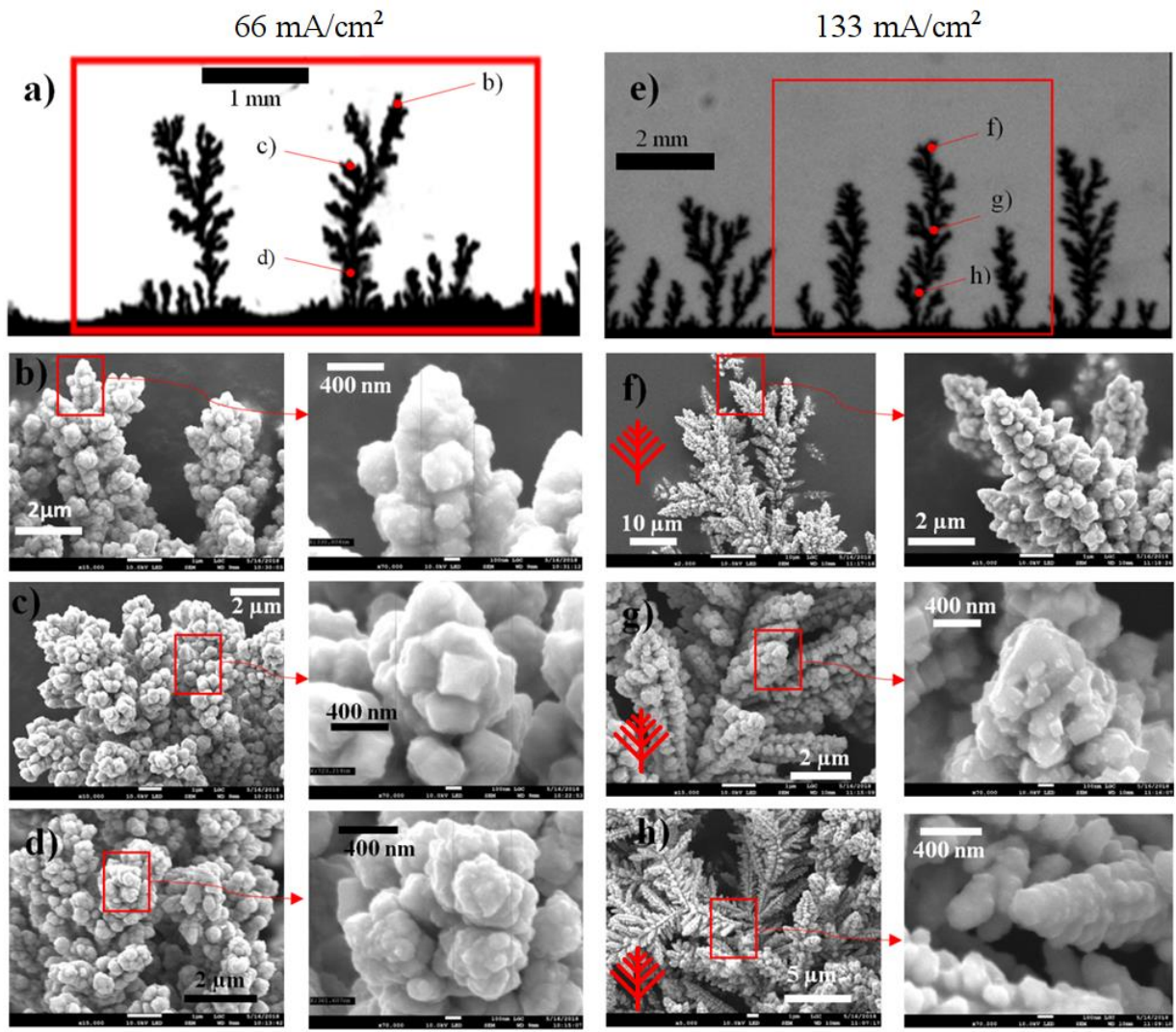

FIG. 5. Optical visualizations (a and e) and SEM observations (b-d and f-h) of copper branches made with $c_0 = 0.5$ M, the applied current density $j$ is 66 mA/cm$^2$ in a-d and 133 mA/cm$^2$ in e-h. Localized SEM observations at the top (b and f), the middle (c and g) and at the bottom (d and h) of the considered branches. The detection of the presence of dendrites is indicated by the inserted sketches.

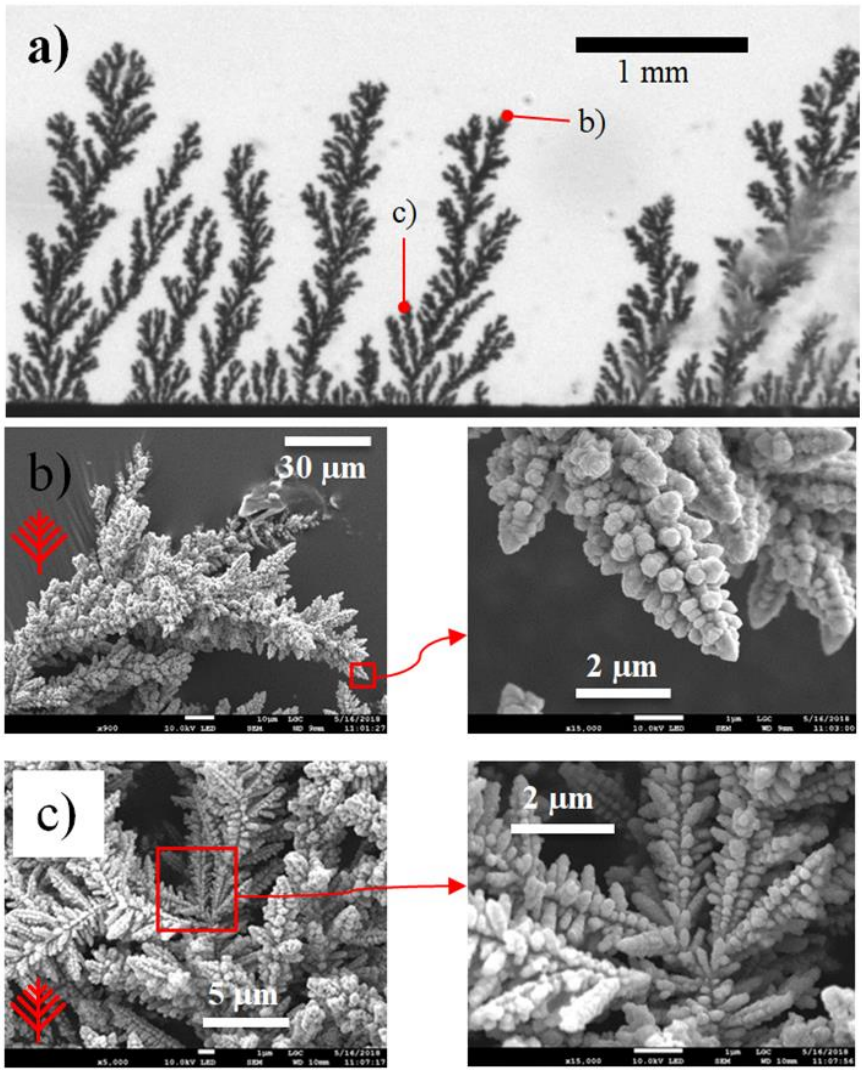

FIG. 6. Optical visualization (a) and SEM observations (b-c) of copper branches made with $j$ = 266 mA/cm$^2$ and $c_0$ = 0.5 M. Localized SEM observations at the top (b) and the middle (c) of the considered branch. The detection of the presence of dendrites is indicated by the inserted sketches.

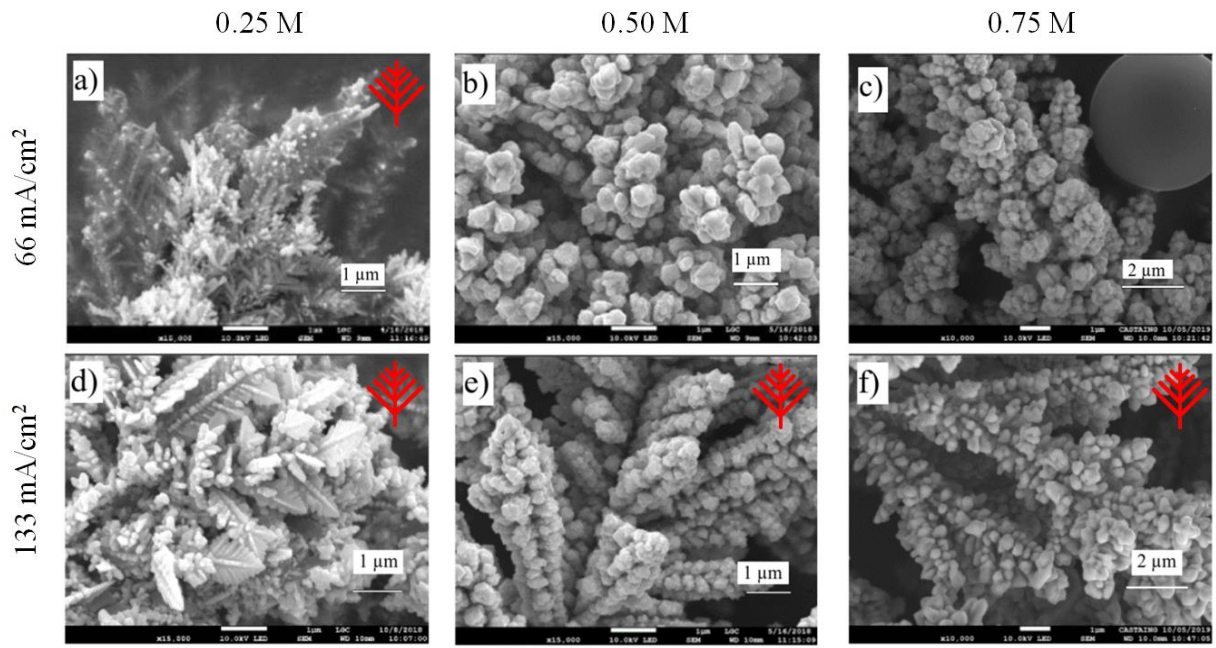

FIG. 7. SEM observations of copper branches, in their middle, for three concentrations $c_0$ of 0.25, 0.50 and 0.75 M, and for two applied current densities $j$ of 66 mA/cm$^2$ and 133 mA/cm$^2$. The detection of the presence of dendrites is indicated by the inserted sketches.

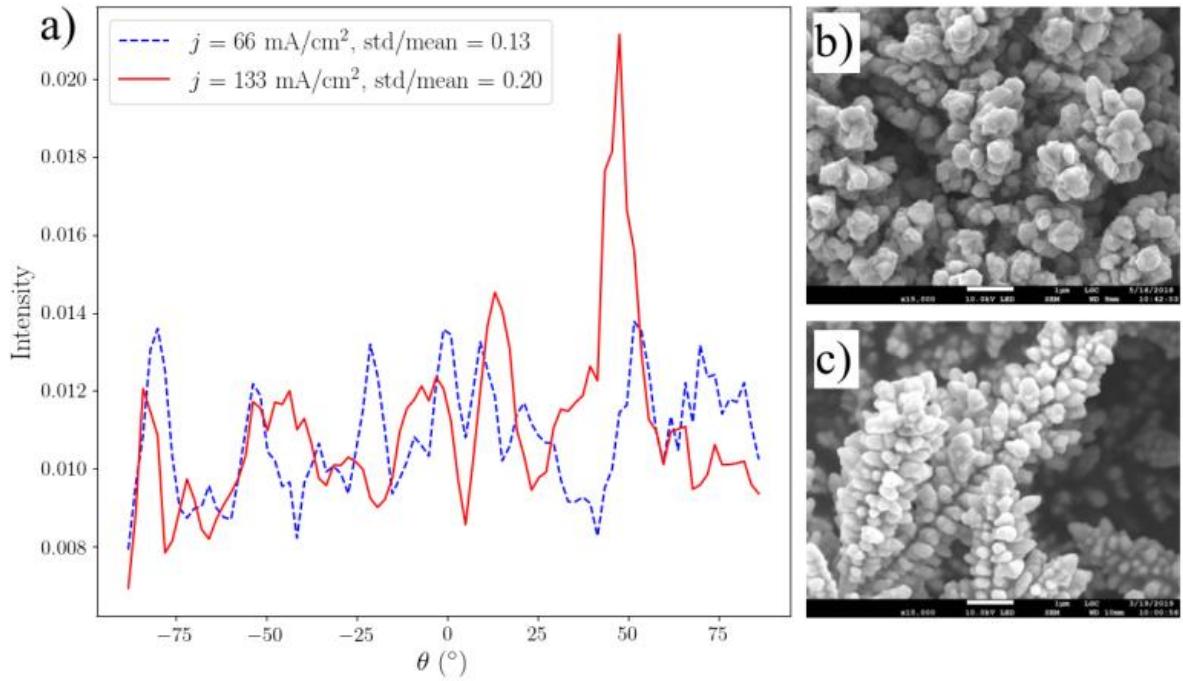

FIG. 8. a) Directionality plots (from the Fiji software), corresponding to the indicator of preferential orientations as a function of the orientation $\theta$ relative to the image frame ($\theta = 0$ corresponds to the East direction and the direction of rotation is counterclockwise), for 66 mA/cm$^2$ and 133 mA/cm$^2$ ([CuSO$_4$] = 0.5 M); for each plot, the ratio between the standard deviation and the mean value (std/mean) is provided in the legend; the SEM images used are shown in b) (66 mA/cm$^2$) and c) (133 mA/cm$^2$). In a), the peak at ~45° corresponds to the long dendrite shown in c).

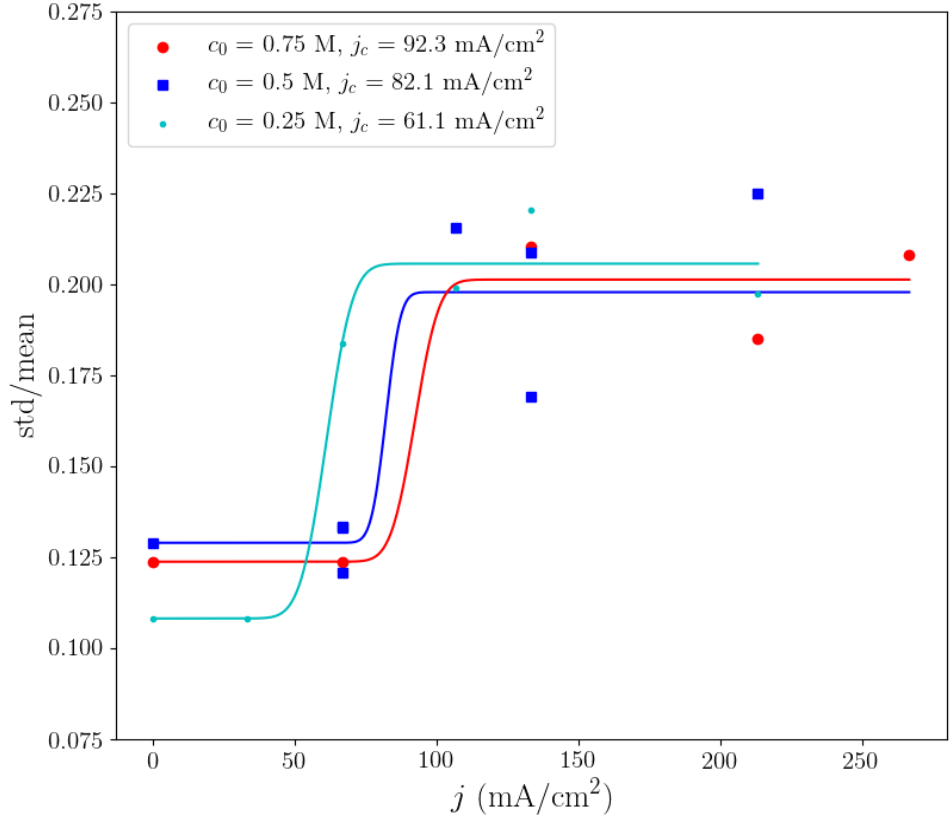

FIG. 9. The dendrite descriptor (ratio between the standard deviation and the mean value of the indicator of preferential orientations) as a function of the current density for various $CuSO_4$ concentrations $c_0$. Data are fitted by an erf function which is plotted for each concentration as a solid line in the same color as the corresponding symbol; the determined critical current density $j_c$ is given in the legend.

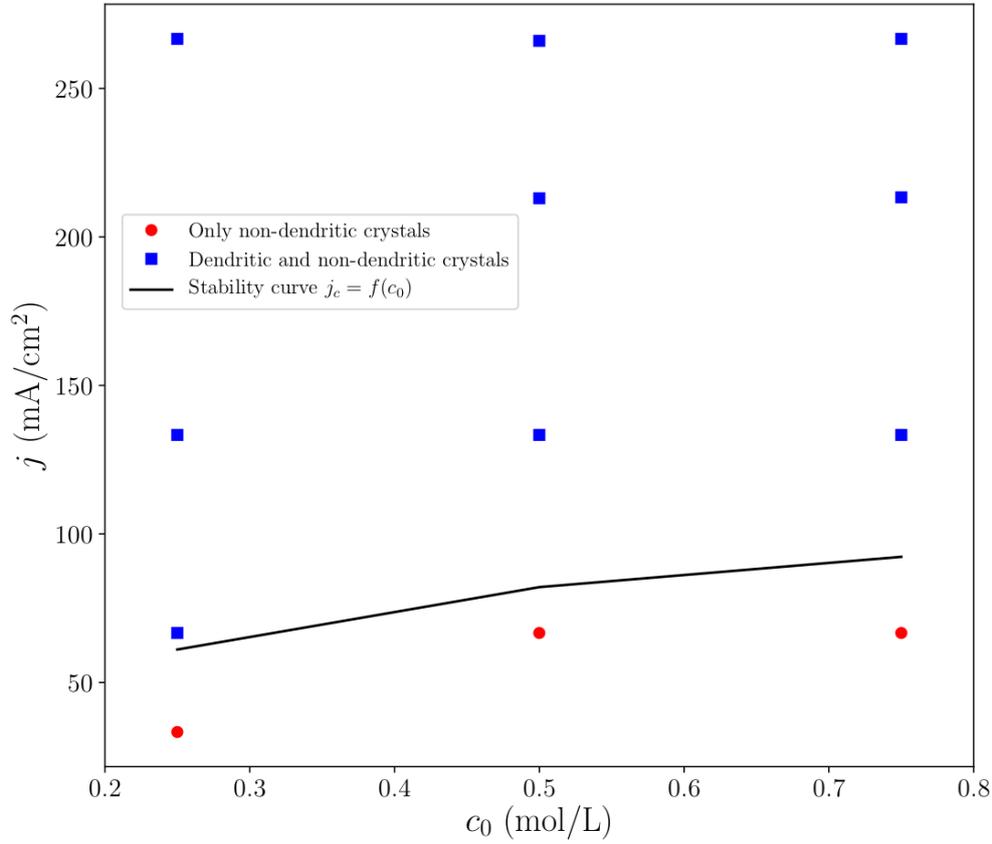

FIG. 10. Stability diagram of the shape of the crystals constituting the branches in the form of the plot of the applied current density $j$ as a function of the concentration $c_0$. Filled red circles and blue squares correspond to cases where respectively only non-dendritic crystals and dendritic, with possibly non-dendritic, crystals are observed. A stability curve (black line) is plotted by joining the points $j_c = f(c_0)$ obtained from image processing of the SEM images.

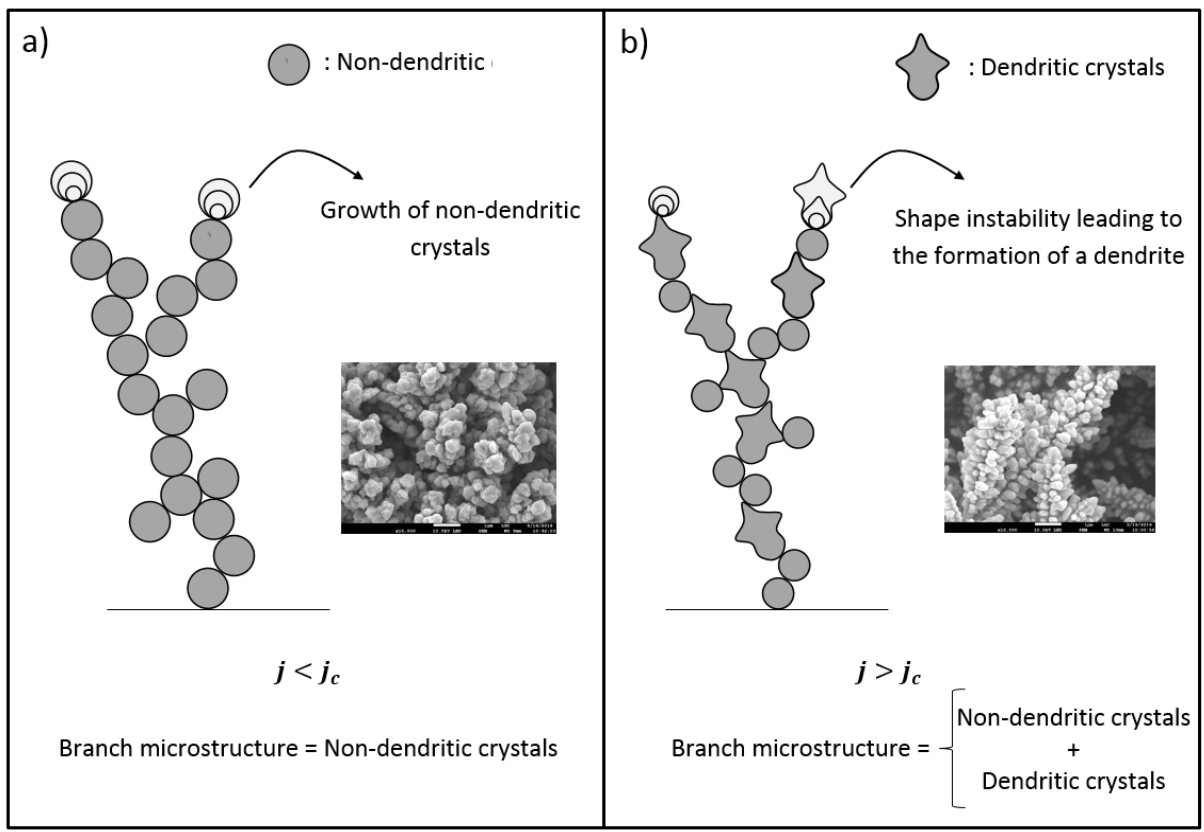

FIG. 11. Sketch of branch microstructure for $j < j_c$ in a) and $j > j_c$ in b). The SEM images are the same as in Fig. 8.

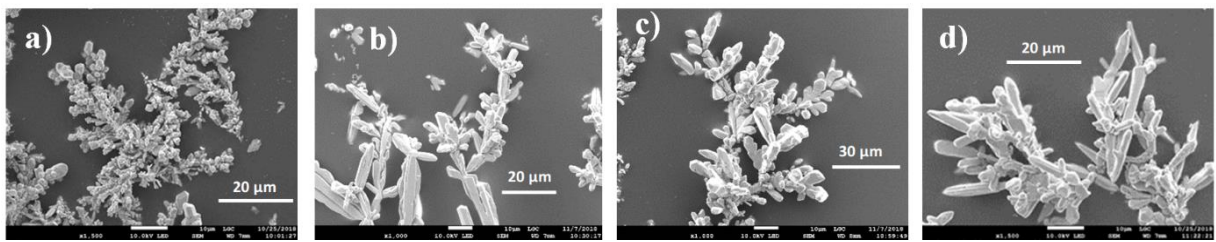

FIG.12. Microstructure of silver ramified branches (top of the branches) observed by SEM and made at the following operating conditions: [AgNO$_3$] = $c_0$ = 0.25 M, and $j$ = 33 mA/cm$^2$ (a), $j$ = 66 mA/cm$^2$ (b), $j$ = 133 mA/cm$^2$ (c) and $j$ = 266 mA/cm$^2$ (d); note that the magnification is the same for each case (except c)).

# IV. ON THE ONSET OF DENDRITIC GROWTH ON THE SCALE OF THE MICROSTUCTURE

The transition from the nucleation/growth of non-dendritic crystals to the growth of dendrites, when $j$ exceeds $j_c$, is theoretically considered here. It is widely accepted that the onset of dendritic growth is fundamentally due to a shape instability during the growth of an initially stable shape growing material (whatever the initial shape), as described in several references [28–30]. The shape instability is also responsible for the transient emission of side branching during the growth of the dendrite tip. The shape instability occurs when the amplification rate of infinitesimal protrusions is higher than their damping rate. The Mullins & Sekerka model [25], adapted to diffusion-limited growth, isotropic materials and fast kinetics, usually describes this mechanism. The amplification or the damping of shape fluctuations is dictated by the balance between destabilizing effects, generally a "point effect of diffusion", and stabilizing effects which are surface energy and kinetics. If the growth is slow, the stabilizing effects have time to damp any eventual deviations/protrusions to the stable shape and consequently the particle is stable keeping its equilibrium shape (non-dendritic particle). While if the growth is fast, the amplification rates of the protrusions are higher than the damping rates and the particle is unstable; this results in the formation of a dendrite.

As shown in the experimental part, dendrites arise from instability of initially non-dendritic crystals. Consequently, the shape stability of a growing crystal, idealized as spherical, is considered. We consider the shape stability threshold formulated in term of a threshold crystal size $R_t$. If the size of the growing crystals $R(t)$ stays lower than $R_t$, during their growth duration $T$, no dendritic growth can occur. Dendritic growth is considered possible if $R(T) = R_g > R_t$.

Concerning the spherical shape considered in this modeling, this is an idealization but this is not expected to affect the roles of the main physical effects. Indeed, as it could be seen on the SEM images of the microstructure, the non-dendritic crystals are not spherical. Therefore, there is an anisotropic effect as expected for the deposited materials (also shown by the dendrites themselves). But, as discussed jut before, the onset of dendritic growth, even for anisotropic material, is also fundamentally due to a shape instability [29,31]. This latter is usually described by a Mullins & Sekerka like model in which orientation-dependent surface energy and kinetic coefficient are taken into account. In this case, the situation is more complex since the threshold depends on the orientation [31]. In the present work, it is not relevant and necessary to take into account all the complexity brought by anisotropies. Because, even if this would be done, the derived threshold would show the same trend (as a function of the operating parameters) as if an isotropic case would be considered. This is due to the fact that the competition between the stabilizing and destabilizing effects remains intrinsically the same regardless of the taking into account of the anisotropy and therefore of the shape of the growing crystal. Instead, an isotropic situation is considered, and so a spherical shape is considered, but the transport of cations around the growing crystal is specifically considered.

In the following, from the Mullins & Sekerka shape stability analysis, the developed modeling aims to derive a relationship for $R_t$. The modeling of the growth and shape of the resulting dendrites (that would require the taking into account of material anisotropies) is not attempted. As it stands, the Mullins & Sekerka shape stability analysis (adapted to crystallization, condensation, etc.) is not directly adapted to the electrochemical situation. Firstly, because species transport involves at least two fields (electrolyte concentration and electric potential) instead of just one. Secondly, because the driving force is the overpotential instead of the oversaturation. Consequently, a specific shape stability analysis is derived here.

A growing crystal at the top of a branch is assimilated to an initially spherical metal particle which grows, under the action of a cathodic polarization, in a stagnant solution of the metal salt (electrolyte), Fig. 13.

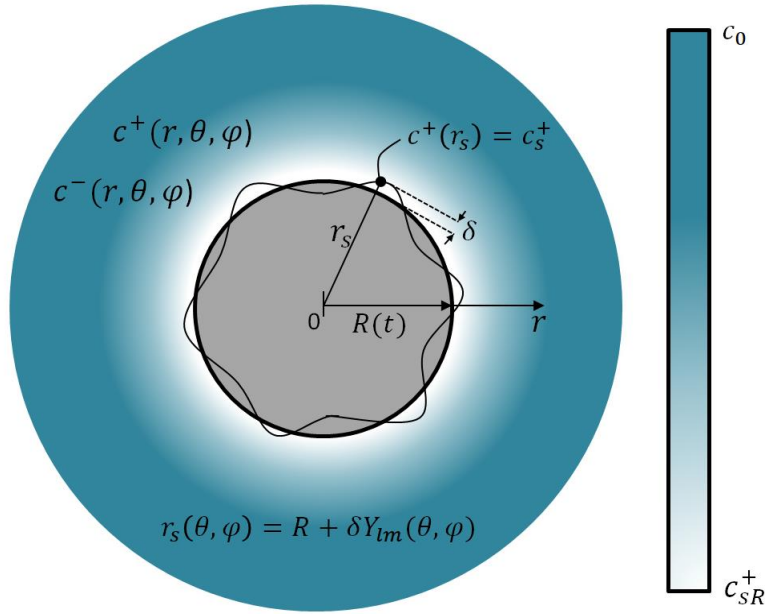

FIG. 13. Sketch of the electrochemical growth of a spherical crystal (dark grey), surrounded by an electrically neutral and diffusional region (color map), and subject to shape deformations whose amplitude is measured by $\delta(t)$.

To enable the derivation of the stability analysis (using spherical harmonics, see below), the particle is considered as a single particle. The relevant scalar fields (concentrations and electric potential) are thus subject to spherical symmetry, Fig. 13. Eventual departure from the electroneutrality is not considered in this application of the Mullins & Sekerka model. Consequently, the cation concentration $c^+$ is equal to the anion concentration $c^-$ in the solution (the electrolyte is considered symmetric).

Even if there are diffusion and migration of ions (no supporting electrolyte), the transport problem can be mathematically reduced to a simple diffusion process (transport in a binary electrolyte) [5,33]. The particle growth can therefore be considered as diffusion-limited. Furthermore, this diffusion process is quasi-stationary because the concentration field adapts sufficiently rapidly to the change in particle size. This is directly related to the strong difference in metal density between the particle (solid metal) and the liquid (dissolved metal salt), see also the discussion in [25]. As a consequence, the Laplace equation is satisfied in the liquid:

$$\nabla^2 c^+ = 0. \tag{1}$$

The cation concentration at the particle surface $c_s^+$ depends on both the overpotential $\eta$ and the local curvature $K$ [34,35]:

$$\eta(t) = \frac{RT}{zF}\log\left(\frac{c_s^+(t)}{c_0}\right) - \frac{\gamma_{PL}V_m}{zF}K(t), \tag{2}$$

where R is the ideal gas constant, F the Faraday's constant, T the temperature, $\gamma_{PL}$ is the free surface energy of the interface between the particle and the liquid, $V_m$ the molar volume, $z$ the ion valence, and $K$ the local curvature of the interface. In the right hand side of this latter equation, the first term corresponds to the concentration overpotential and the second one is a correction to account for the curvature of the interface (Gibbs-Thomson effect); for a spherical particle of radius $R$, $K = 2/R$. In accordance with the diffusion-limited growth assumption, the activation overpotential, related to the electrochemical kinetics, is neglected.

After linearization assuming small $K$ in Eq. 2, $c_s^+$ depends on shape deformations according to the following relation:

$$c_s^+ = c_{sf}^+(1 + \Gamma K), \tag{3}$$

where $c_{sf}^+ = c_0\exp\left(\frac{zF}{RT}\eta(t)\right)$ is the equilibrium concentration of the cations at the a flat metal surface and $\Gamma = \frac{\gamma_{PL}V_m}{RT}$ the capillary length.

As in classical shape stability analyses, the deformations are modeled with spherical harmonics $Y_{lm}(\theta,\varphi)$ of degree $l$ and mode $m$ ($\theta$ and $\varphi$ being the angular spherical coordinates). The location of the particle surface $r_s$ is given by (Fig. 13):

$$r_s(t,\theta,\varphi) = R(t) + \delta(t)Y_{lm}(\theta,\varphi), \tag{4}$$

where $\delta$ is the deformation amplitude. We aim to find from which particle radius $R$, the amplification rate $\dot{\delta}/\delta$ becomes positive.

To the first order in $\delta$, the solution of the Laplace equation (Eq. 1) satisfying the boundary condition Eq. 3 is given by (by identification with Eq. 7 in [25]):

$$c^+(r,\theta,\varphi) = \frac{(c_{sf}^+ - c_0)R + 2c_{sf}^+\Gamma}{r} + c_0 + \frac{(c_{sf}^+ - c_0)R^l + c_{sf}^+\Gamma R^{l-1}l(l+1)}{r^{l+1}}\delta Y_{lm}(\theta,\varphi)$$

. (5)

The conservation of the deposited metal at the particle surface leads to:

$$v\rho = \frac{D}{1-t_c}\frac{\partial c^+}{\partial r}\bigg|_{r_s} + \frac{v}{1-t_c}c^+|_{r_s},$$ (6)

where $\rho$ is the metal density (mol/m³), $D$ the mean diffusion coefficient defined by $D = (u_c D_a + u_a D_c)/(u_c + u_a)$ (where $D_a$, $D_c$, $u_a$ and $u_c$ are the diffusion coefficients and the mobilities of respectively anions and cations), $t_c$ the transference number of the cations [33] and $v = dr_s/dt$ the normal velocity of the surface [25].

Combining Eq. 5 and 6, the interface velocity expresses as (Eq.8 in [25]):

$$\dot{R} + \dot{\delta} = \frac{D}{\rho(1-t_c) - c_{sR}^+}\left[\frac{c_0 - c_{sR}^+}{R} + \frac{\delta Y_{lm}}{R}\left((l-1)\frac{c_0 - c_{sf}^+}{R} - \frac{\Gamma}{R^2}c_{sf}^+(l(l+1)^2 - 4)\right)\right]$$

, (7)

where $c_{sR}^+ = c_{sf}^+(1 + 2\Gamma/R)$. In the right hand side of Eq. 7, the term in square brackets is exactly the same as in [25]. Mullins & Sekerka show that shape instability requires that the deformations grow faster than the particle expansion, $(\dot{\delta}/\delta)/(\dot{R}/R) > 1$, and $l \geq 3$. This leads to the definition of the threshold radius $R_t$:

$$R_t = 42\Gamma\frac{c_{sf}^+}{c_0 - c_{sf}^+}.$$ (8)

Shape deformations appear if $R$ exceeds $R_t$. Contrary to usual phase changes (crystallization, condensation, etc.), where the surface concentration is constant (equilibrium saturation), here, $c_{sf}^+$ changes during the particle growth because of its dependence on the overpotential, $c_{sf}^+ =$ $c_0 \exp((zF/RT)\eta)$ (Eq. 3). Consequently, the stability condition ($R_t$) changes during the particle growth.

The periodic temporal evolution of $\eta$ is sketched in Fig. 14b and explained here. According to Fleury's nucleation/growth model, just after the nucleation of a new particle, the incoming current is concentrated on its surface to make it grow at a constant growth rate (the previous particle does no longer growth, Fig. 14a); the particle radius thus follows a one-third power law on a time period of $T$, Fig. 14a.

Initially, the particle size $R(0)$ is very low, typically ~1 nm (corresponding to overpotential ~100 mV). This induces a so high interface velocity ($\propto R^{-2}$) that both space charge and electrolyte depletion do not have time to develop (as initially postulated by Fleury [22]) and this prevents $c_s^+/c_0$ to reach small values. At the early stage of the particle growth, the Gibbs-Thomson overpotential then overcomes the concentration overpotential $\eta(t) \approx -\frac{\gamma_{PL} V_m}{zF} \frac{1}{R(t)}$. This induces a fast decrease of $\eta$ at the beginning of a cycle, Fig. 14b.

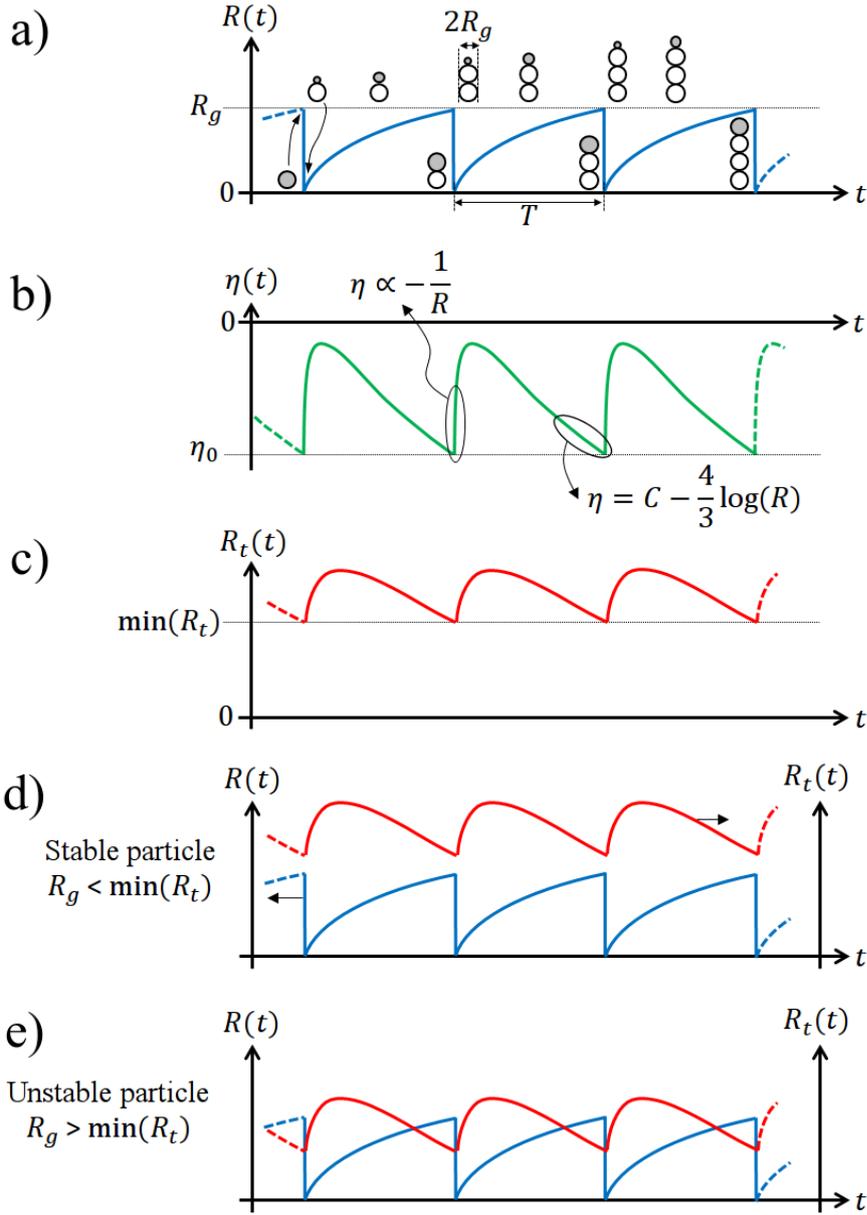

FIG. 14. Sketches of the periodic temporal evolution of radius of a growing particle $R$ (a), overpotential $\eta$ (b) and threshold radius for the onset of shape instability assuming a diffusion-limited growth $R_t$ given by Eq. 8 (c). In a), successive nucleation/growth events are sketched, the top particle, colored in gray, is the particle which is growing. The $R(t)$ and $R_t(t)$ signals are plotted in the same graph for both cases stable particle (d) and unstable particle (e).

For longer times, up to the end of particle growth, the interface velocity is well lowered (because the particle is now large, $R \rightarrow R_g \sim 100$ nm, Fig. 5 and 6) and both electrolyte depletion and space charge have time to develop [22]. This leads to $c_s^+/c_0 \ll 1$ and the overpotential now corresponds to the concentration overpotential $\eta(t) \approx \frac{RT}{zF} \log\left(\frac{c_s^+(t)}{c_0}\right)$. As it is shown below, for this regime, $c_s^+$ is proportional to $j_g(t)^{2/3}$ (see also Eq. A21 with A19 and A20), with $j_g(t)$ the current density on the growing particle surface. The assumption of constant growth rate induces $j_g \propto R^{-2}$ and consequently $\eta(t) \approx C - \frac{4}{3}\log(R(t))$ (where $C$ is a constant) as sketched in Fig. 13b. When $\eta$ reaches a critical value $\eta_0 = \eta(0) = \eta(T)$, a new particle nucleates and grows immediately on the one that was growing and a new cycle starts, Fig. 13a. Note that the temporal signal of $\eta$ is very similar to the temporal signal of the surface electric field shown in [22] (Fig. 4).

The corresponding signal of $R_t$ is plotted in Fig. 13c. On a growth cycle, the instability is favored at the end of particle growth. This differs from usual situations where $R_t$ is constant. As sketched in Fig. 13d-e, the stability condition for a growing particle to be stable on a full cycle is $R_g < \min(R_t)$.

The estimation of $\min(R_t)$ requires $\eta_0$ (Eq. 8) which can be derived at the end of a particle growth. For these long growth times, the electrolyte depletion at the electrode surface induces a divergent electric field (assuming electroneutrality). By considering the corresponding Poisson-Nernst-Planck problem, assuming a stationary and one-dimensional system, Chazalviel [2] showed the very high electric field at the electrode surface leads to the formation of a space charge region. The electrolyte phase consists of two regions [2,22]: in the vicinity of the electrode (particle/crystal) surface, on a certain small thickness $x_I$ (~ 100 nm), the solution is not electrically neutral, $c^+ \gg c^-$, this is the charge space region, while beyond $x_I$, the solution is neutral ($c^+ = c^-$). In the space charge region, both cation concentration and electric potential profiles can be derived according to Chazalviel's model [2]. This gives access to $c_s^+(t = T)$ and therefore to $\eta_0 \approx \frac{RT}{zF} \log\left(\frac{c_s^+(T)}{c_0}\right)$.

Note that the use of Chazalviel's model requires the system to be stationary. This is actually the case for long growth times for which particles grow so slowly that mass transport, across the space charge by both diffusion and migration, can be considered as quasi-stationary. This is justified by the estimation of the characteristic growth time $T = 2R_g/v_g$, the characteristic diffusion time $T_d = x_I^2/D_c$ and the characteristic migration time $T_m = x_I/(zF|E_s|D_c/(RT))$. By estimating $x_I$ and $|E_s|$ from the stationary and one-dimensional Chazalviel's theory (Eq. 23, 25 and 27 in [2] and considering a potential drop across the space charge region, $|E_s|x_I$, of 1 V, Appendix B) and considering the typical encountered ranges for $R_g$ [150, 550 nm] and $v_g$ [1, 20 µm/s] (for $j \in$ [33, 100 mA/cm²] and $c_0 \in$ [0.50, 0.75 M], Appendix C), $T_d/T$ and $T_m/T$ are lower than 0.07%. This greatly simplifies the modeling contrary to the beginning of a particle growth for which the problem is transient and requires numerical simulation [22].

Nevertheless, $x_I$, computed from Chazalviel's model, which is one-dimensional, ($x_I$ ~100 nm), is of the same order of magnitude as the size of the particles (~100 nm) at the transition ($j = j_c$). Consequently, we expect that $x_I$ depends on the particle radius.

Here, the Chazalviel's model is revisited taking into account a spherical geometry (Appendix A). For a spherical electrode (particle) of radius $R$ subject to a current density $j_g$, from the theoretical framework of Chazalviel [2], we obtain the following profiles across the space charge layer for both electric field and cation concentration:

$$\phi(x_I + R) - \phi(r) = R\sqrt{\frac{2j_g}{3zFu_c\varepsilon\varepsilon_0}} \int_r^{x_I+R} \frac{\sqrt{(x_I + R)^3 - x^3}}{x^2} dx$$

, (9)

$$c^+(r) = \left[\frac{2(zF)^3 u_c}{3j_g\varepsilon\varepsilon_0}\left(\frac{(x_I + R)^3 - r^3}{R^2}\right)\right]^{-1/2}$$

, (10)

for $R + x_I \geq r \geq R$ and where $\varepsilon_0$ is the vacuum permittivity, $\varepsilon$ the relative permittivity of water and $r$ the radial coordinate. The potential drop across the space charge layer $\delta V$ is obtained from Eq. 9 ($\delta V = \phi(x_I + R) - \phi(R)$):

$$\delta V = R\sqrt{\frac{2j_g}{3z\mathrm{F}u_c\varepsilon\varepsilon_0}}\int_R^{x_I+R}\frac{\sqrt{(x_I+R)^3-x^3}}{x^2}dx$$

, (11)

$\delta V$ has been measured and it is around 1 V for each case (Appendix B). This is in agreement with similar measurements [24,36].

The link between $j_g$ and the operating parameters ($j$ and $c_0$) can be established from Fleury's nucleation/growth model. An important statement of this model is the link between microstructure and macrostructure (pattern) trough the relation [22]:

$$T = \frac{2R_g}{v_g},\qquad(12)$$

where $v_g$ is the average growth velocity of the branches on the scale of the branch pattern. Note that Eq. 12 is obtained by neglecting the size of the nucleus $R(0)$ in front of $R(T) = R_g$. As experimentally verified several times [5,11,37] and established theoretically [2], $v_g$ corresponds to the anion velocity:

$$v_g = -z\mathrm{F}u_a E,\qquad(13)$$

where $E$ is the mean electric field. From the expression of the current density inside the electrolyte, where there are no concentration gradients, $j = -z^2\mathrm{F}^2(u_a + u_c)c_0 E$, Eq. 13 can be given in a form in which the operating parameters ($j$ and $c_0$) appear explicitly:

$$v_g = \frac{1-t_c}{z\mathrm{F}}\frac{j}{c_0}.\qquad(14)$$

Note that Eq. 14 can also be obtained by considering, in one dimension, the stationary advection-diffusion of the electrolyte above the growth front [5]. The measured values of $v_g$ for the present experiments follow well Eq. 14 (Appendix C).

According to Fleury's nucleation/growth model [22], the particle growth rate is constant and the Faraday's law leads to:

$$\frac{I_g}{zF} = \frac{d}{dt}\left(\frac{4}{3}\pi\rho R^3\right), \tag{15}$$

Where $I_g = (4\pi R^2)j_g$ is the incoming current on one growing particle. After integration of Eq. 15, coupled with Eq. 12 and 14, we obtain a relation for the current density at the end of the growth phase $j_g(T)$:

$$j_g(T) = \frac{1}{6}zF\rho v_g. \tag{16}$$

Eq. 16 is an important relation because it gives a direct access to the current density on the particle surface at the end of growth, as a function of the operating parameters through $v_g$ (Eq. 14) whatever the arrangement of the branches on the pattern scale.

For each set of input parameters ($j$, $c_0$, $R = R_g$), the concentration $c_{sf}^+$, which is equal to the surface concentration $c_s^+(T) = c^+(R_g)$, can therefore be estimated combining Eq. 10, 11, 14 and 16. Note that $x_I$ is determined by numerically solving Eq. 11, jointly using a non-linear solver and a numerical integration; $x_I$ is found to have a very weak dependence on $R$ unlike $c_s^+(T)$ (Appendix A). The estimated values of $c_{sf}^+$ are so low (maximum value of $10^{-3}$ M, Appendix A, Fig. A1c) that the growing particles are expected to be always unstable. The prefactor of $\delta$ in the second member of the right hand side of Eq. 7 is always positive. The damping of shape deformations ($\dot{\delta}/\delta < 0$, in Eq. 7) would require prohibitive values of $l$ (>100) for the typical particle size $R$ of 100 nm: $R_g \gg \min(R_t)$. This is not consistent with the experiments which clearly show that there is a stable growth regime of the crystals.

Furthermore, it is interesting to analyze the variation of $\min(R_t)$ with $j$ and $c_0$. According to Eq. 8, $\min(R_t) \approx 42\, \Gamma c_s^+(T)/c_0$. The value of $c_s^+(T)$ results from the specific interaction between both cation concentration and electric potential, in the space charge layer. Indeed, across the space charge layer, the current flow is mainly ensured by the migration of cations (Appendix A, Eq. A15), $((zF)^2 u_c)\delta V c_s^+(T)/x_I \approx j_g(T)$. Since the potential drop $\delta V$ is constant and that $j_g(T) \propto j/c_0$ (Eq. 14 and 16), $c_s^+(T)/x_I$ is found to be proportional to $j/c_0$. By considering the Poisson equation (Eq. A14), $\delta V/x_I^2 \approx (zF/(\varepsilon\varepsilon_0))c_s^+(T)$, the following power laws are obtained: $x_I \propto (j/c_0)^{-1/3}$ and $c_s^+(T) \propto (j/c_0)^{2/3}$. If $j/c_0$ increases (i.e. $v_g$ increases), this induces an increase of the electric field ($\approx \delta V/x_I$) which is limited, to avoid divergence, by the enrichment in cations, $c_s^+(T)$ increases.

Consequently, $\min(R_t)$ increases with $j$ ($\min(R_t) \propto j^{2/3}$) that is also not consistent because this would favor stability instead of instability when $j$ is increased. Additionally, $\min(R_t)$ decreases with the concentration $c_0$ ($\min(R_t) \propto c_0^{-5/3}$). This would suggest that $j_c$ decreases with $c_0$. This trend is also not correlated with the experimental results (Fig. 10).

These differences show that the *standard* shape stability analysis, adapted from Mullins & Sekerka model, is not able to describe the appearance of dendrites on the level of branch microstructure. This probably comes from the too simplified model of mass transport which considers only diffusion. Indeed, in the vicinity of the particle surface, due to the presence of the space charge layer, the mass transport problem can no longer be assimilated to a diffusion situation. In the space charge layer, the migration overcomes the diffusion of cations ( [2] and Appendix A) and this affects the stability condition. Even if a full derivation is required (this will be done in another work), we could predict the main effect of the space charge on the shape stability.

By considering the limitation by migration of the cations, Eq. 6 becomes:

$$v = \frac{zu_c \mathrm{F}}{\rho - c^+|_{r_s}} \left[ c^+ \frac{\partial \phi}{\partial r} \right]_{r_s}. \tag{17}$$

In the space charge layer, the amplitude of the variation of $\phi$ is $\sim \delta V$, on a length of about $x_I$ and therefore $\left.\frac{\partial \phi}{\partial r}\right|_{r_s} \approx \delta V/x_I$. As in diffusion-limited cases, there is therefore a destabilizing point effect but induced by the electric potential across a shell (gradient zone) which the thickness is $x_I$; this differs significantly from diffusion-limited cases where the shell stays conform to the particle during its growth $\left.\frac{\partial c^+}{\partial r}\right|_{r_s} \approx c_0/R$.

During the growth of a particle, $x_I \propto j_g^{-1/3}$ (Eq. A19), and therefore $x_I \propto t^{2/9}$. For short times, this power law induces a faster increase for $x_I$ than for $R \propto t^{1/3}$ (shell thickness for diffusion-limited cases).

For long times, $t \to T$, $x_I$ reaches typical values around 100 nm and it varies rather slowly with time ($t^{2/9}$). This prevents the amplification of shape deformations ($\delta \ll x_I$) as long as the particle size $R(t)$ is lower than $\sim x_I(T) \sim 100$ nm; the particle grows inside a shell with an almost constant thickness. On the contrary, for a diffusion situation, the point-like effect is likely to appear earlier, for smaller particles, and this leads to a low instability threshold as obtained with Eq. 8 (where the oversaturation $(c_0 - c_{sf}^+)/c_{sf}^+$ is typically high in such an electrochemical situation). The taking into account of a space charge must therefore lead to a positive instability threshold which the order of magnitude is the same as the one of $x_I(T) \sim 100$ nm. This latter is compatible with the actual size of the particles encountered here $R_g \in [150, 550$ nm$]$. Furthermore, since $x_I(T) \propto (j/c_0)^{-1/3}$, this suggests that the onset of shape instability, and thus the presence of dendrites, is favored when $j$ is increased and when $c_0$ is decreased. These predicted trends are now consistent with the experiments (Fig. 10).

### IV. Conclusion

It is experimentally shown that the transition from non-adherent ramified branches (nucleation/growth regime) towards dendrites (growth regime) appears on the scale of the crystals constituting the ramified branches. These non-dendritic crystals undergo a shape instability leading to the formation of dendrites when $j$ exceeds a critical applied current density $j_c$. This latter increases with the electrolyte concentration $c_0$.

The fact that the crystals become unstable when $j$ increases could appear usual (similarly to other crystallization or condensation phenomena) but this is not so simple. Notably because both the crystal size and the oversaturation (Eq. 8 and A21) decrease with the ratio $j/c_0$ that generally prevents shape instability and thus dendritic growth. According to the standard Mullins & Sekerka shape stability analysis (Eq. 8), unlike the experiment results (Fig. 10), these trends suggest an enhanced shape stability when $j/c_0$ increases. This discrepancy is due to the periodic formation of a space charge layer (which avoids the divergence of the electric field, a specificity of the present situation) which strongly modifies the transport problem, compared to classic (diffusion-limited) crystallization or condensation situations.

By considering a space charge layer around a growing crystal, the right trends are qualitatively obtained. It was previously demonstrated that the space charge region plays a major role on the formation of ramified branches (onset of branch growth [2] and re-nucleation process [22]). Here, the obtained results suggest the space charge region also plays an unsuspected role in:

- stabilization of the shape of non-dendritic crystals constituting ramified branches (the dendritic growth is delayed)
- onset of dendritic growth on the scale of the growing crystals.

This shows, once again [22], the originality of this growth phenomenon compared to other solidification or condensation situations.

The obtained improved knowledge on the transition between non-dendritic and dendritic regimes will favor the development of an alternative synthesis of metal nanomaterial based on the exploitation of the nanostructure of the branches.

This study focuses only on the onset of dendritic growth. Many questions remain open especially beyond the dendritic transition, when the branch microstructure is mixed (non-dendritic and dendritic crystals). For example, in the case of copper (as here) and iron ( [38,39]), why the size of the dendrites is always so low that, on the macroscale, the pattern remains ramified ? While for zinc, very large dendrites are obtained [7]. There should be an interaction between re-nucleation process and dendrite growth (beyond shape instability) governed by material (anisotropic) properties.

# APPENDIX A: DERIVATION OF THE RELATIONS OF THE SPACE CHARGE REGION FOR A SPHERICAL PARTICLE

The electrochemical growth of a spherical particle of radius $R(t)$ is considered. The derivation is carried out for the end of the growth ($t \to T$, $R(t) \to R_g$) when the system can be considered quasi-stationary (see section IV-A). In the liquid, $r \geq R(t)$, the concentration of cations and anions and the electric potential satisfy the transport equations:

$$zu_c F \nabla \cdot (c^+ \nabla \phi) + D_c \nabla^2 c^+ = 0, \tag{A1}$$

$$-zu_a F \nabla \cdot (c^- \nabla \phi) + D_a \nabla^2 c^- = 0, \tag{A2}$$

as well as Poisson equation:

$$\nabla^2 \phi = -\frac{F}{\varepsilon \varepsilon_0}(zc^+ - zc^-). \tag{A3}$$

There are no advection terms since they are typically negligible for such small particles (the Péclet number $\sim v_g R_g / D$ is lower than 0.01). Using spherical coordinates, the relations A1-3 become:

$$D_c \frac{d}{dr}\left(r^2 \frac{dc^+}{dr}\right) + zu_c F \frac{d}{dr}\left(r^2 c^+ \frac{d\phi}{dr}\right) = 0, \tag{A4}$$

$$D_a \frac{d}{dr}\left(r^2 \frac{dc^-}{dr}\right) - zu_a F \frac{d}{dr}\left(r^2 c^- \frac{d\phi}{dr}\right) = 0, \tag{A5}$$

$$\frac{1}{r^2}\frac{d}{dr}\left(r^2 \frac{d\phi}{dr}\right) = -\frac{F}{\varepsilon \varepsilon_0}(zc^+ - zc^-). \tag{A6}$$

Integrating Eq. A4 between $R$ and $r \geq R$, considering the total mass flux of cations is $4\pi R^2(j_g/(zF))$ at the particle surface, leads to:

$$D_c \frac{dc^+}{dr} + zu_c F c^+ \frac{d\phi}{dr} = \frac{j_g}{zF}\left(\frac{R}{r}\right)^2. \tag{A7}$$

By doing the same for the anions (Eq. A5), considering the total mass flux of anions is 0 at the particle surface, leads to:

$$D_a \frac{dc^-}{dr} - zu_a Fc^- \frac{d\phi}{dr} = 0. \tag{A8}$$

Integrating Eq. A6 between $r \geq R$ and a distance $L'$ being sufficiently high such that the electrolyte concentration $\approx c_0$, we obtain:

$$c^-(r) = c_0 \exp\left(-\frac{zu_a F}{D_a}(\phi(L') - \phi(r))\right). \tag{A9}$$

Combining Eq. A7 and A8, assuming electroneutrality ($c^- = c^+$) and using $u_c = D_c/RT$ and $u_a = D_a/RT$, we obtain:

$$2D_c \frac{dc^+}{dr} = \frac{j_g}{zF}\left(\frac{R}{r}\right)^2. \tag{A10}$$

Following the theoretical model of Chazalviel [2], the thickness of the charge region $x_I$ is defined such that, for $r > R + x_I$, the electroneutrality applies, and, for $R \leq r < R + x_I$, the electroneutrality does not apply. By integrating Eq. A10 between $R + x_I$ and $r$ the cation (and anion) concentration field is given by:

$$c^+(r) = \frac{j_g}{zF}\left(\frac{R}{r}\right)^2 \frac{1}{2D_c}\left(\frac{R^2}{R+x_I} - \frac{R^2}{r}\right), \tag{A11}$$

where, $c^+(R + x_I)$ is neglected in front of $c^+(r)$ in this neutral region (as in [2]). By combining Eq. A8 and A10, the gradient of electric potential, in the neutral region, is given by:

$$\frac{d\phi}{dr} = \frac{D_a}{zu_a F}\left(\frac{1}{\frac{r^2}{R+x_I}-r}\right). \tag{A12}$$

By integrating this latter equation between $r$ ($> R + x_I$) and $L'$, we obtain for the electric potential:

$$\phi(r) = \phi(L') - \frac{D_a}{zu_a F}\left[\log\left(\frac{L'-(R+x_I)}{r-(R+x_I)}\right) - \log\left(\frac{L'}{R+x_I}\right)\right]$$

. (A13)

Eq. A13 shows that $\phi(r \to R + x_I) \to -\infty$, and from Eq. A9, $c^-(r \to R + x_I) \to 0$. Since $\phi(r)$ is expected to be a monotonic function, $c^- \to 0$ in the space charge region (see also concentration profiles obtained numerically [2]).

In the space charge region, Poisson equation becomes:

$$\frac{1}{r^2}\frac{d}{dr}\left(r^2 \frac{d\phi}{dr}\right) = -\frac{F}{\varepsilon\varepsilon_0}zc^+. \tag{A14}$$

Because of the high electric field in the space charge region (Eq. A8), the transport of cations is mainly due to migration and Eq. A7 becomes:

$$zu_c F c^+ \frac{d\phi}{dr} = \frac{j_g}{zF}\left(\frac{R}{r}\right)^2. \tag{A15}$$

Combining Eq. A14 and A15, after integration between $r$ and $R + x_I$ (and considering $c^+(R + x_I) \gg c^+(r)$), leads to:

$$c^+(r) = \left[\frac{2}{3}\frac{(zF)^3 u_c}{\varepsilon\varepsilon_0 j_g}\left(\frac{(R+x_I)^3 - r^3}{R^2}\right)\right]^{-1/2}. \tag{A16}$$

From Eq. A15, after integration, the electric potential is given by:

$$\phi(r) = \phi(x_I + R) - R\sqrt{\frac{2j_g}{3zFu_c\varepsilon\varepsilon_0}}\int_r^{x_I+R}\frac{\sqrt{(x_I+R)^3-x^3}}{x^2}dx. \tag{A17}$$

From the latter equation, introducing $\delta V = \phi(x_I + R) - \phi(R)$ the potential drop across the space charge, the thickness of the space charge layer $x_I$ can be deduced from:

$$\delta V = R\sqrt{\frac{2j_g}{3zFu_c\varepsilon\varepsilon_0}}\int_R^{x_I+R}\frac{\sqrt{(x_I+R)^3-x^3}}{x^2}dx. \tag{A18}$$

The values obtained for $x_I$ and $c^+(R = R_g) = c_s^+(T)$, for this spherical case, are compared to the prediction of the one-dimensional model of Chazalviel (Eq. 27 and 25 in [2]]):

$$x_I^{1D} = \left(\frac{3}{2}\right)^{2/3}\delta V^{2/3}\left(\frac{\varepsilon\varepsilon_0 zFu_c}{2j_g}\right)^{1/3}, \tag{A19}$$

$$c_s^{+1D} = \left(\frac{\varepsilon\varepsilon_0}{2(zF)^3 u_c} \frac{j_g}{x_I^{1D}}\right)^{1/2}. \tag{A20}$$

$x_I$ is found to stay close to $x_I^{1D}$ whatever the ratios $j/c_0$ and $x_I/R$ (for the ranges encountered here), Fig. A1a. As expected, $x_I \to x_I^{1D}$ for $x_I/R \to 0$, Fig. A1b.

The same trend is obtained for $c_s^+(T)/c_s^{+1D}$ but $c_s^+(T)$ is significantly lower than $c_s^{+1D}$ for high values of $x_I/R$, Fig. A1c. A useful observation is that the curves $c_s^+(T)/c_s^{+1D}$, as a function of $x_I/R$, coincide for all values of $j/c_0$, Fig. A1d. This enables a separation of the variables in the relation providing $c_s^+(T)$ for the spherical case:

$$c_s^+(T) = f(x_I/R)\left[c_s^{+1D}(j, c_0)\right], \tag{A21}$$

where the indeterminate function $f(x_I/R)$ can be roughly considered as an exponential function, $f(x) = \exp(-0.336x)$, Fig. A1d.

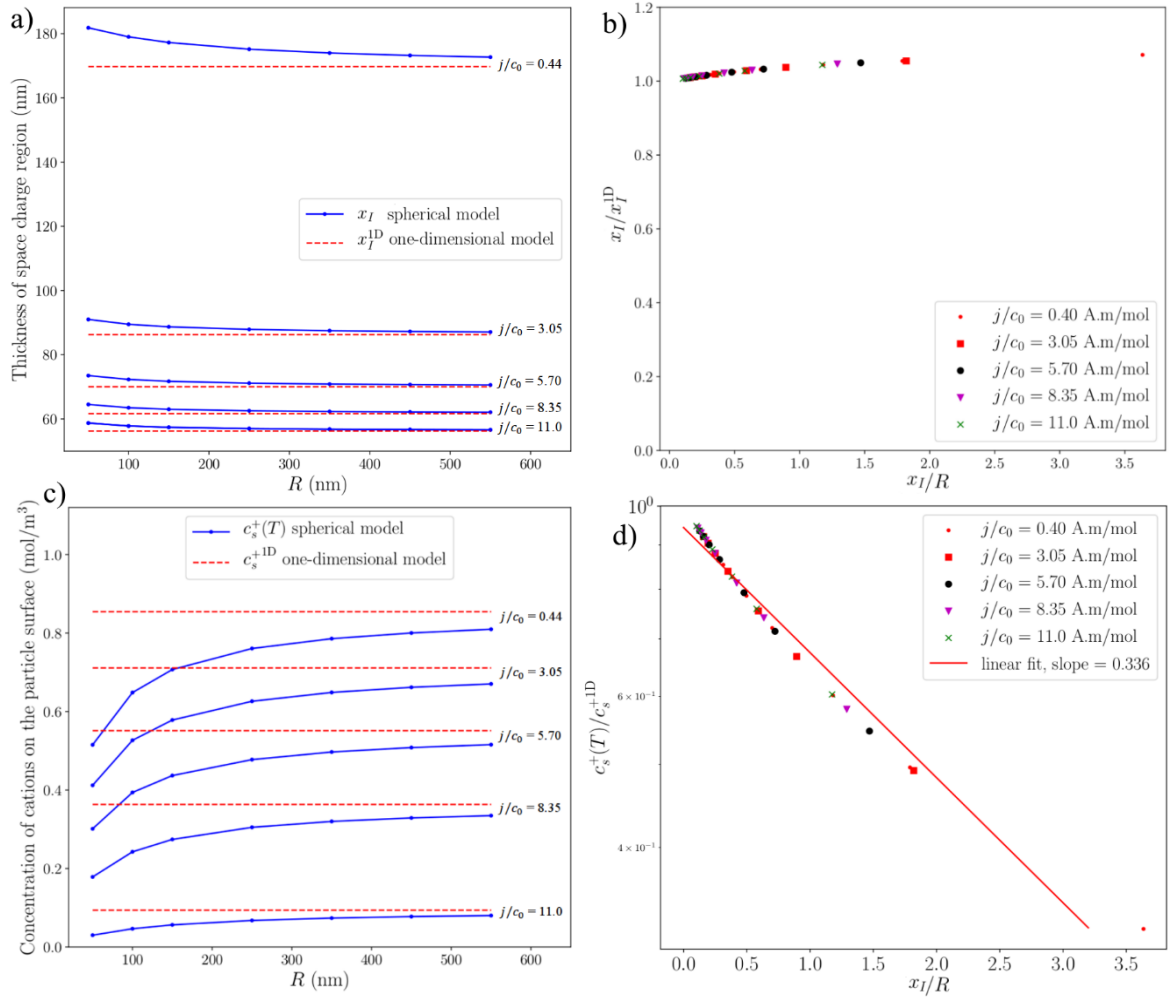

FIG. A1. Comparison between spherical and one-dimensional models for the prediction of both space charge region/layer thickness (a,b) and cation surface concentration (c,d). Space charge region thicknesses ($x_I$ for the spherical model, $x_I^{1D}$ for the one-dimensional model) (a) and the ratio $x_I/x_I^{1D}$ (b) as a function of respectively the particle radius $R$ and the ratio $x_I/R$, for several values of the ratio $j/c_0$. Cation surface concentrations ($c_s^+(T)$ for the spherical model, $c_s^{+1D}$ for the one-dimensional model) (c) and the ratio $c_s^+(T)/c_s^{+1D}$ (d) as a function of respectively the particle radius $R$ and the ratio $x_I/R$, for several values of $j/c_0$ (given in A.m/mol).

# APPENDIX B: ESTIMATION OF THE POTENTIAL DROP ACROSS THE SPACE CHARGE REGION

During a galvanostatic electrolysis, the cell tension $\Delta V$ evolves as shown in Fig. B1a for the particular case $c_0 = 0.5$ M, $j = 133$ mA/cm$^2$. Just after the application of the current, there is a decrease of $\Delta V$ up to ~5 s. This initial variation is related to a capacitive effect on a rather large time scale compared to usual electrochemical situations (~ 1 ms); de Bruyn explained this effect by the influence of natural convection [36].

At the same time, an increasing part of the applied current (all the current from the moment when $\Delta V$ stops to decrease with $t$) is used to convert the cations into metal at the cathode surface (electrochemical reduction). Due to the non-renewal of the electrolyte (stagnant solution) there is a depletion region close to the initially flat cathode surface. At the surface of the cathode, the concentration tends towards ~0. The instant when the concentration reaches ~0 (Sand's time) marks the onset of branch growth [40]. At the same time, cell tension rises (partly because of the increase in the resistivity, itself induced by electrolyte depletion) up to a maximum value, Fig. B1a. The time at the peak is close to Sand's time and it also marks the beginning of branch growth [36,40].

Next, $\Delta V$ decreases linearly with time because the growth front (cathode) gets closer to the anode and the corresponding inter-electrode distance decreases linearly with time (as shown and discussed in several previous works [11,36], see also Appendix C). By considering that the branches are a perfect electrical conductor, $\Delta V$ is the sum of several contributions:

$$\Delta V = |\eta| + \eta_a + \delta V' + \delta V_{conc} + \delta V_{ohm}, \tag{B1}$$

where:

- $\eta_a$ is the anodic overpotential
- $\delta V_{ohm}$ the potential drop where the concentration field is uniform (the enriched anodic region is not considered), ahead the region of concentration gradients (Fig. B2)

- $\delta V_{conc}$ the potential drop, where there are concentration gradients and electroneutrality of the solution (typically on a length $L_D$ related to the average growth velocity by $L_D = D/v_g$), between the region with uniform concentration field and the region where the fields (concentration and potential) cannot be considered as one-dimensional (due to the branch microstructure)
- $\delta V'$ is the remaining potential drop across the space charge region and a small part of the neutral region, Fig. B2.

By considering the system as one dimensional, from a distance $H$ (~$R_g \in [150, 550\text{ nm}]$) above the top of the branches (Fig. B2), the two last terms in the right hand side of Eq. B1 are given by:

$$\delta V_{conc} = \int_{\bar{x}_f+H}^{\bar{x}_f+H+L_D} d\phi, \tag{B2}$$

$$\delta V_{ohm} = \rho_e(c_0)j(L - \bar{x}_f - H - L_D), \tag{B3}$$

where $\bar{x}_f$ is the average location of the growth front (Fig. B2) and $\rho_e(c_0)$ the electric resistivity of the solution. Assuming that the overpotentials are negligible compared to the other contributions (this is generally satisfied, and even comforted for fast kinetics as considered here, since the overpotentials fall in the range ~10-100 mV) as well as $\delta V_{conc}$, Eq. B1 takes the simplified form:

$$\Delta V \approx \delta V' + \delta V_{ohm}. \tag{B4}$$

$\delta V_{conc}$ has been neglected compared to $\delta V_{ohm}$, as in previous works [24,36]. This can be justified because $\delta V_{conc}$ slowly diverges when the concentration at the growth front tends towards ~0. Indeed, in this region, the dependence of the electric potential on the local concentration can be deduced from the general expression of the electric current inside the electrolyte which is given here by ($c = c^+ = c^-$):

$$j = z^2 F^2 (u_a + u_c) c(x) \frac{\partial \phi}{\partial x} + zF(D_c - D_a) \frac{\partial c}{\partial x}. \tag{B5}$$

This latter equation can be simplified by neglecting the diffusion term (this is reasonable, first because the diffusion coefficients are expected to be close, and second because this term does not contain diverging terms contrary to the migration term):

$$j \approx z^2 F^2 (u_a + u_c) c(x) \frac{\partial \phi}{\partial x}. \tag{B6}$$

In this neutral region, the concentration profile satisfies the diffusion equation [33]:

$$\frac{\partial c}{\partial t} = D \frac{\partial c}{\partial x^2}, \tag{B7}$$

which can be converted to its stationary form using the coordinate relative to the front $x' = x - v_g t$:

$$\frac{d^2 c}{dx'^2} + L_D^{-1} \frac{dc}{dx'} = 0 \tag{B8}$$

A solution of this latter equation is:

$$\frac{c(x') - c_0}{c^* - c_0} = \exp\left(-\frac{x' - H}{L_D}\right), \tag{B9}$$

where $c^*$ is the electrolyte concentration at $x' = H$ (Fig. B2). By integrating Eq. B6 between $\bar{x}_f + H$ and $\bar{x}_f + H + L_D$, using Eq. B9, we obtain:

$$\delta V_{conc} = \frac{j}{z^2 F^2 (u_a + u_c)} \frac{L_D}{c_0} \log\left(1 + (\exp(1) - 1) \frac{c_0}{c^*}\right). \tag{B10}$$

Since $L_D = D/v_g$ and $v_g \propto j/c_0$ (Eq. 16), according to Eq. B10, $\delta V_{conc}$ does not directly depend on $j$ nor $c_0$. Eq. B10 shows that the divergence of $\delta V_{conc}$ with $1/c^*$ is slight because of the logarithmic dependence. The value of the electrolyte depletion $c_0/c^*$, and its trend with both $j$ and $c_0$, can be estimated by taking the value of $c^+(R) = c_s^+(T)$, given by Eq. A20, for $c^*$ ($= c(\bar{x}_f + H)$); note that $c_0/c^*$ and $\delta V_{conc}$ are therefore overestimated. We find that $\delta V_{conc}$ decreases with $j$ and for the worst case (0.75 M, $j = 33$ mA/cm²), $\delta V_{conc} < 0.093$ V. This rather low upper value, and the fact that $\delta V_{conc}$ decreases with $j$, (see below and Fig. B1b) allows this term to be neglected.

Considering a cell length $L$ well higher than $L_D$ and $\bar{x}_f$ (and $x_I$), $\delta V_{ohm} \approx \rho_e(c_0)jL$, and its value can be estimated from the value of $\Delta V$ just after the initial capacitive phase (for $t < \sim 5$ s in Fig. B1a) when the concentration profile has not changed much yet (this value is taken as the minimum value on the plot $\Delta V$ as a function of time). By subtracting this last value from the maximum value of $\Delta V$ (at the peak), we measure $\delta V'$. $\delta V'$ thus determined actually correspond to the actual space charge region at the initially flat cathode surface ($\delta V'$ = potential drop across the space charge region for a one-dimensional system for which $H = 0$). Since, after the peak, there is no discontinuity of the cell tension and that it decreases linearly with time, this suggests that $\delta V'$ remains almost the same even if the electroactive surface changes significantly (flat surface → top of the ramified branches, $H$ can no longer be considered as 0). We deduce that this estimation of $\delta V'$ gives access to $\delta V$.

Overall, we obtain values for $\delta V$ of about 1 V. The same value has already been reported for the same system [24]. As already observed in the work of De Bruyn 1997 [36], we observe a slow decrease in $\delta V$ with the applied current density $j$ (note that this trend is not expected to be due to not taking into account $\delta V_{conc}$ since this latter would induce the opposite trend as discussed just above).

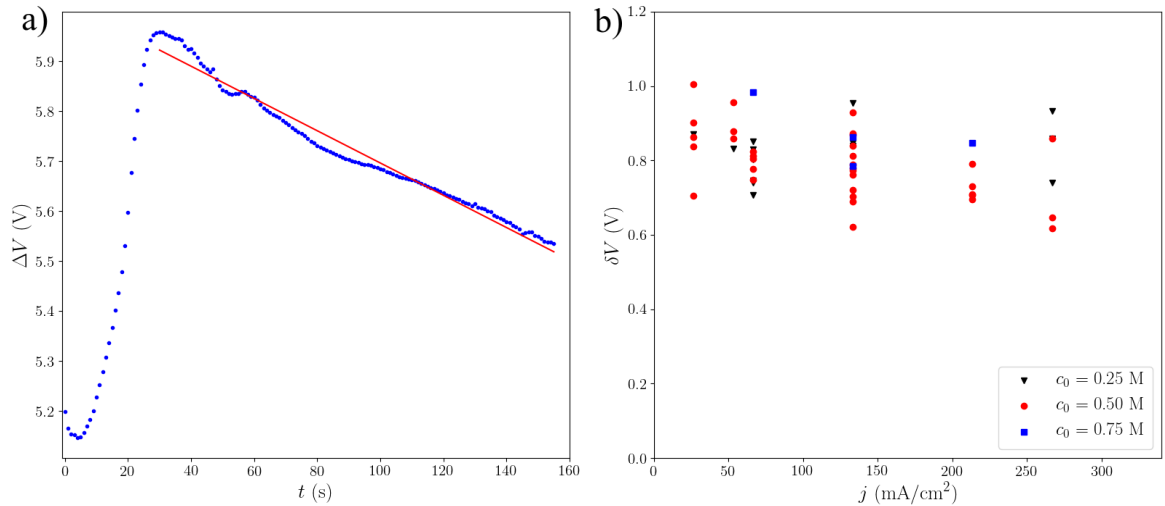

FIG. B1. a) The cell tension $\Delta V$ as a function of time for $c_0 = 0.5$ M and $j = 133$ mA/cm², the red line (corresponding to the linear fit of the decreasing part of the curve on the right of the peak) highlights the linear decrease of $\Delta V$ with $t$. b) The estimated potential drop across the space charge region $\delta V$ as a function of the applied current density $j$ and for several concentrations $c_0$.

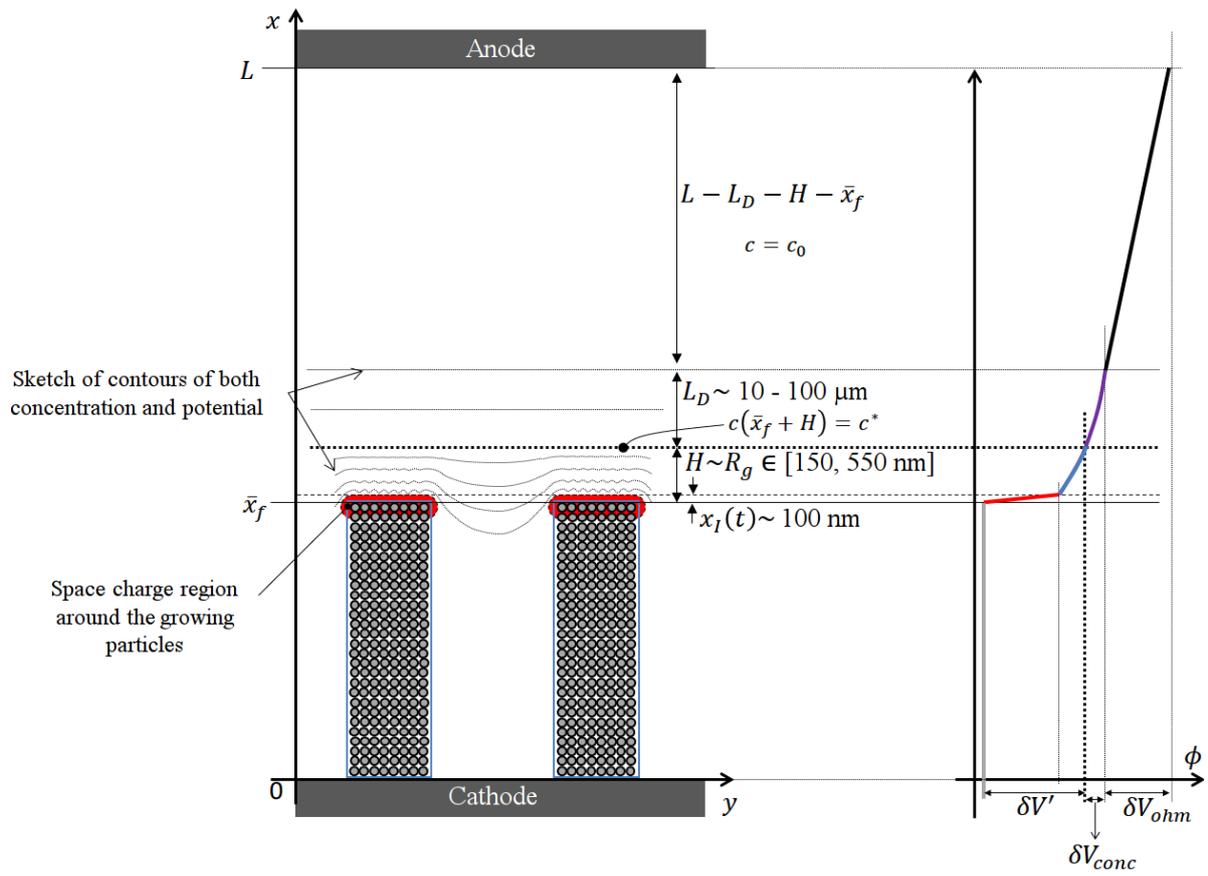

FIG. B2. Sketch of the electric potential profile across the electrochemical cell.

# APPENDIX C: MEASUREMENT OF THE AVERAGE GROWTH VELOCITY

The average growth velocity $v_g$ is defined as $d\bar{x}_f/dt$, where $\bar{x}_f$ is the average location of the top of growing branches along the $x$ coordinate (along the axis perpendicular to the initial surface of the cathode and directed towards the anode), Fig. C1. Here, $\bar{x}_f$ is determined from the half height of the wave of the density profile $\rho_p(x)$ obtained by image processing (Fig. C1b): $\rho_p(x) = \left(\int_0^W A(x,y)dy\right)/W$, where $W$ is the width of the image, $A = 1$ where there is metal and $A = 0$ elsewhere and $y$ is the coordinate along the axis parallel to the initial electrode surfaces, Fig. C1a. In Fig. C1b, $\rho_p$ profiles are shown for different times during branch growth for $c_0 = 0.5$ M and $j = 266$ mA/cm². The corresponding time evolution of $\bar{x}_f$ is shown in Fig. C1b.

At short times (for $t < $ ~40 s), $\bar{x}_f = 0$, branch growth is not observed. However, copper is deposited during this phase but without re-nucleation process (mainly growth); the visualization tool used does not allow the corresponding displacement of the cathode surface to be measured. This feature is usually observed during branch formation by galvanostatic electrolysis and is due to the initial depletion of the electrolyte in the vicinity of the cathode surface. This depletion is due to the non-renewal of the electroactive species (the cations) in this *stagnant* solution. When the electrolyte concentration reaches ~0 at the electrode surface, either the growth front moves forward (formation of the branches) or another electrochemical reaction (generally the reduction of the free protons or of the water) takes place or both co-exist, in order to ensure the flow of the imposed electric current [5]. In the case of copper and silver in the present study, we observe only the formation of branches for the investigated range of the operating parameters.

The average growth velocity is obtained from the slope of the plot of $\bar{x}_f$ as a function of time (in the growing part of the curve) from a linear fit, as shown in the inset of Fig. C1b.

As already obtained for DB copper deposits [11], the dependence of $v_g$ with both $j$ and $c_0$ matches the predicted growth velocity, based on the velocity of the anions $v_a$, (Eq. 14), $v_g = (1 - t_c)/z\text{F}(j/c_0)$ as shown in Fig. C2. From a linear fit of these latter data, we obtained for $t_c$ a value of 0.393 (a value in accordance with other works on copper branches [5]).

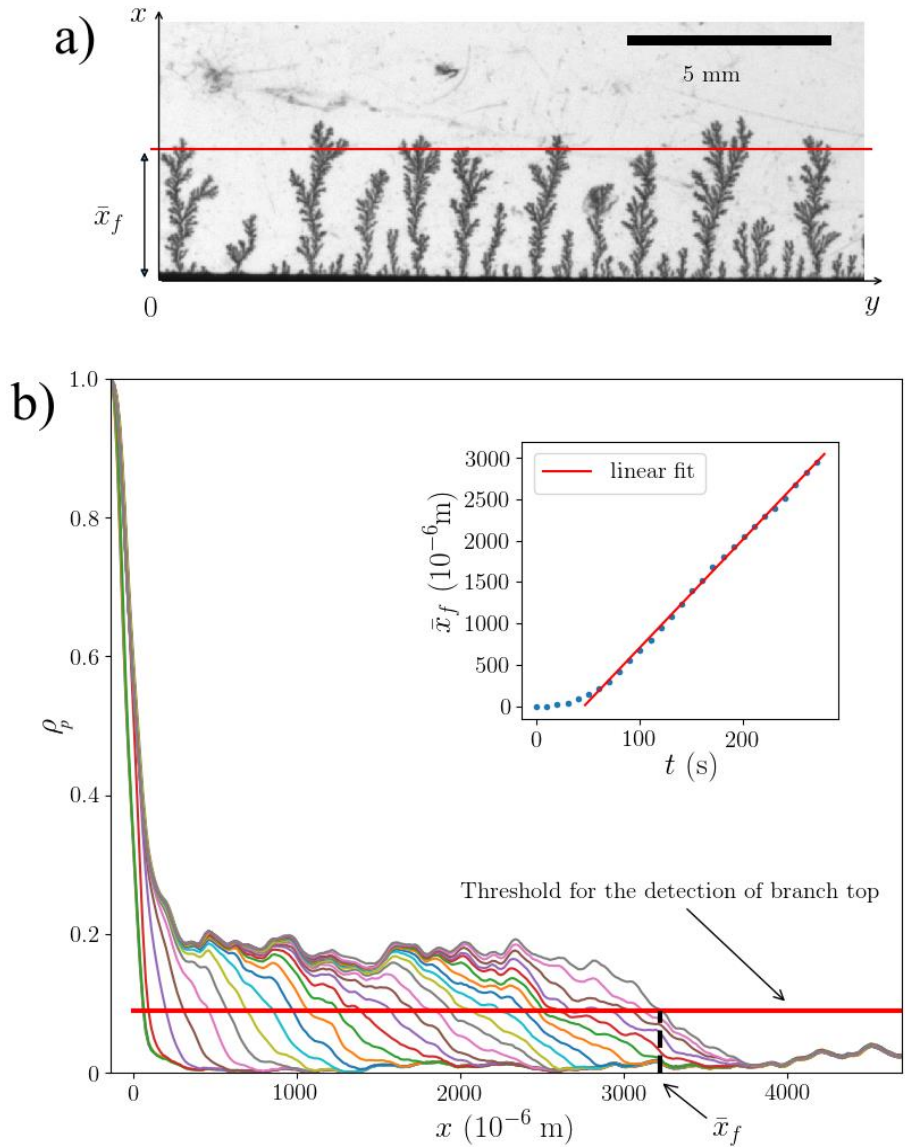

FIG. C1. a) Optical visualization of a ramified copper deposit obtained with $c_0 = 0.5$ M, $j = 266$ mA/cm$^2$ after an electrolysis duration of 280 s, the red line indicates the location of the top of growing branches, $\bar{x}_f$, detected from the last density profile shown in b). b) The corresponding $\rho_p$ profiles (see the text) at several times (there is a gap of 10 s between each curve), the straight and horizontal red line corresponds to the applied threshold for the measurement of $\bar{x}_f$; the inset shows the temporal evolution of $\bar{x}_f$ as a function of time for the case shown in a).

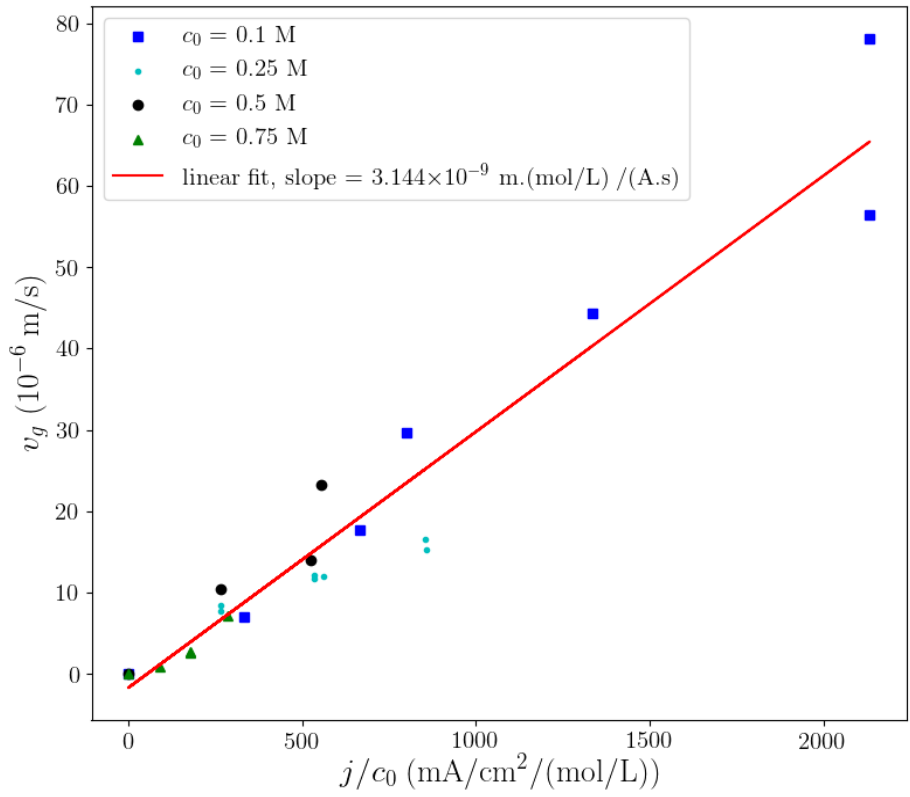

FIG. C2. The average growth velocity $v_g$ as a function of the ratio $j/c_0$ for all of the investigated concentrations; the red line corresponds to the linear fit of all the points (the slope = $3.144 \times 10^{-9}$ m.(mol/L)/(A.s)).